\documentclass[fleqn,usenatbib]{mnras}

\usepackage{newtxtext,newtxmath}

\usepackage[T1]{fontenc}

\DeclareRobustCommand{\VAN}[3]{#2}
\let\VANthebibliography\thebibliography
\def\thebibliography{\DeclareRobustCommand{\VAN}[3]{##3}\VANthebibliography}

\usepackage{graphicx}	
\usepackage{amsmath}	
\usepackage{amssymb}	
\usepackage{subfig}
\usepackage{comment, bm}
\graphicspath{{./}{figures/}}

\usepackage{ulem}

\title[CCNN]{Concat Convolutional Neural Network for Pulsar Candidate Selection}

\author[Zeng et al.]{
Qingguo Zeng,$^{1}$
Xiangru Li,$^{2}$\thanks{Correspondence author's E-mail: xiangru.li@gmain.com (X.Li)},
Haitao Lin$^{1}$
\\
$^{1}$School of Mathematical Sciences,South China Normal University, No. 55 West of Yat-sen Avenue, Guangzhou 510631, China\\
$^{2}$School of Computer Science, South China Normal University, Guangzhou 510631, China
}

\date{Accepted 2020 March 30. Received 2020 March 29; in original form 2020 February 21}

\pubyear{2020}

\begin{document}
\label{firstpage}
\pagerange{\pageref{firstpage}--\pageref{lastpage}}
\maketitle

\begin{abstract}
Pulsar searching is essential for the scientific research in the field of physics and astrophysics. As the development of the radio telescope, the exploding volume and it growth speed of candidates growth have brought about several challenges. Therefore, there is an urgent demand for developing an automatic, accurate and efficient pulsar candidate selection method. To meet this need, this work designed a Concat Convolutional Neural Network (CCNN) to identify the candidates collected from the Five-hundred-meter Aperture Spherical Telescope (FAST) data. The CCNN extracts some ``pulsar-like'' patterns from the diagnostic subplots using Convolutional Neural Network (CNN) and combines these CNN features by a concatenate layer. Therefore, the CCNN is an end-to-end learning model without any need for any intermediate labels, which makes CCNN suitable for the online learning pipeline of pulsar candidate selection. Experimental results on FAST data show that the CCNN outperforms the available state-of-the-art models in similar scenario. It only misses 4 real pulsars out of 326 totally.
\end{abstract}




\section{Introduction}

Pulsars are rapidly rotating, super-dense neutron stars with strong magnetic fields. The rotation of the pulsar causes the beam of electromagnetic radiation field to sweep in and out of our line of sight with an extremely regular period. The theory and observation of the pulsars are of great significance to promote the development of physics and astrophysics, such as the evolution of neutron stars \citep{international1987origin}, the equation of state of dense matter \citep{backer1982millisecond}, verification on general relativity \citep{hulse1975discovery, lyne2004double}, etc. In particular, Pulsar Timing Array (PTA) with dozens of millisecond pulsars can be used to detect and analyse gravitational waves due to their accurate timing properties \citep{van2011placing, manchester2013parkes, demorest2012limits}. Therefore, it is essential to discover new pulsars to excavate their enormous potentials for scientific research.

Ever since Jocelyn Bell Burnell and Antony Hewish observed the first pulsar in 1967 \citep{hewish1968observation}, more than 2700 pulsars have been discovered \citep{manchester2005australia} by the modern radio telescope survey projects, such as  the Parkes multi-beam pulsar survey \citep[PMPS; ][]{manchester2001parkes}, the Pulsar Arecibo L-band Feed Array survey \citep[PALFA;][]{deneva2009arecibo}, low-frequency array (LOFAR) tied-array all-sky survey \citep[LOTASS;][]{coenen2014lofar}, etc. However, astronomers prophesied that the total number of potentially observable pulsars in the Galaxy should be approximately 10 times more than this based on the pulsar population model \citep{faucher2006birth}. To search for more pulsars, some advanced modern radio telescopes will or have been built, such as the Square Kilometre Array \citep[SKA;][]{smits2009pulsar} and Five-hundred-meter Aperture Spherical radio Telescope \citep[FAST;][]{nan2011five}. Specifically, FAST began to be constructed in 2011 and started formal operations on 11 January 2020 \citep{xinhua}. It is expected to discover about 1500 new normal pulsars and 200 millisecond pulsars \citep{yue2012fast}. In practice, the FAST 19-beam drift-scan survey generates more than one million pulsar candidates per night \citep{wang2019pulsar}. However, the proportion of real pulsars among candidates is exceedingly small \citep[approximately one in ten thousand;][]{lyon2013study} due to the presence of radio frequency interference (RFI) and noise. Therefore, it is seldom to select the pulsars among the candidates just by using simple metrics like the signal-to-noise ratio (S/N). Traditionally, pulsar candidates selection is completed through inspecting diagnostic plots of the candidates by human experts, but it is impractical to deal with such extreme volume of candidates in this way. In other words, there exist urgent demands for developing an automatic, accurate and efficient pulsar candidate selection method.

The goal of the pulsar candidate selection is to minimize the retention of the non-pulsar signals without missing pulsar candidates as much as possible, thereby reducing the labor of further observations. In the past few years, a variety of pulsar candidate selection methods have been proposed. Based on the principles of a method, they can be divided into three categories. The first category is of traditional scoring methods. \citet{lee2013peace} ranked the candidates according to their scores which are the linear combinations of six well-designed quality factors. The second category improved the methods by applying machine learning (ML) algorithms to learn how to combine the pre-designed quality factors (usually called features in ML) instead of the artificial combination \citep{eatough2010selection, bates2012high, van2013neutron, morello2014spinn, lyon2016fifty}. In these methods, pulsar candidate selection was served as a binary classification problem. One of the important factors affecting the classification result is the feature design which relies heavily on human experience. An incomprehensive feature design scheme may have a bad effect on the performance of the models. For example, some methods extracted six features just from the pulse profile and dispersion measure (DM) curve. As a result, it is likely to mistakenly identify some RFI candidates as pulsars. These misclassified candidates are often generated by RFI within several frequency channels so that they have the ``pulsar-like'' appearance in both the pulse profile and DM curve. In practice, human experts can identify the pulsars from the candidates just by observing the diagnostic plots successfully. Under this inspiration, the third category attempts to directly utilize the diagnostic plots as the inputs into the model instead of hand-crafted features \citep{zhu2014searching, guo2019pulsar, wangyuanchao2019pulsar, wang2019pulsar}. These methods prompt the model to learn the ``pulsar-like'' patterns from the diagnostic subplots by themselves through data-driven learning. \citet{zhu2014searching} and \citet{wang2019pulsar} proposed a two-layer ensemble model to identify the pulsars. For example, the model in \citet{wang2019pulsar} is composed of five classifiers totally, including two Residual Neural Networks (ResNets), two Support Vector Machines (SVMs) and one Logistic regression (LR). Two ResNets are used to determine whether the time versus phase plot and frequency versus phase plot are ``pulsar-like'', respectively. Two SVMs evaluate how ``pulsar-like'' the pulse profile and DM curve are, respectively. Finally, the LR classifies the candidates based on the output scores from the first four classifiers. The first four classifiers constitute a layer of data processors and this layer is referred to as the first layer. The LR constitutes the second layer which receives the outputs from the first layer. \textbf{However, the label of the first layer (the labels of each diagnostic subplot) may not be in accordance with the candidates' label. In other words, the subplots of some RFI candidates may be the same as those of the pulsars. As a result, we have to manually label whether each of the four subplots is "pulsar-like" individually for every training data, leading to a lot of extra labor.}

In this work, we propose a novel deep learning scheme, Concat Convolutional Neural Network (CCNN), for the pulsar candidate selection based on Convolutional Neural Network (CNN). In this proposed model, a concatenate layer is introduced to replace the second layer of the PICS or PICS-ResNet for overcoming the problem of the non-correspondence between the candidate's nature (the candidate is pulsar or not) and the labels of each diagnostic subplot (the subplot is ``pulsar-like'' or not). In addition, the CCNN extracts the features from four diagnostic subplots only using the CNN: one-dimensional CNN for the pulse profile and DM curve while two-dimensional CNN for the time versus phase plot and frequency versus phase plot. In application, two-dimensional CNN has shown its outstanding ability to deal with image pattern recognition and at the same time, one-dimensional CNN has been proved that it is adept at signal processing and recognition \citep{huang2019signal}. Therefore, the CNN for extracting features rather than the traditional ML models has the great potential to promote the performance of the model to identify the pulsar with the diagnostic subplots as the input into the model. This scheme belongs to an end-to-end learning model. The complex relationship between the target (the identification results of the candidates) and the inputs (four diagnostic subplots) can be directly described by just one single layer without any intermediate processes or intermediate labels. Therefore, the end-to-end learning makes the CCNN suitable for the online learning pipeline of the pulsar candidate selection. By the way, for an online learning pipeline, the newly confirmed candidates can be directly appended to the training dataset to continuously improve the classification accuracy of the model.

The rest of this article is organized as follows: the experimental data and data pre-processing methods are described in the next section. In Section \ref{sec:method}, we presented the components and detailed structure of the CCNN. According to the direction of the concatenate operation, CCNN can be subdivided into Horizontal CCNN (H-CCNN) and Vertical CCNN (V-CCNN). Their performances are investigated and compared with the available methods in Section \ref{sec:Results_Discussion}. We conclude and discuss the future work of pulsar candidate selection for FAST in the final section.

\section{Data}\label{sec:data}
The work is conducted for the Commensal Radio Astronomy FasT Survey \citep[CRAFTS;][]{li2018fast}. CRAFTS is a drfit-scan survey which aims to observe the entire visible sky of the FAST for HI emission and search for the new pulsars utilizing the FAST L-band Array of 19-beams \citep[FLAN;][]{zhang2019status}. The early observation data with labels \citep{wang2019pulsar} for pulsar searching from CRAFTS has been public on \url{https://github.com/dzuwhf/FAST_label_data}. This work uses this data set to train and test our model.

\subsection{The information of the dataset}
The dataset has been split into the training set and test set. The training set consists of 837 real pulsars and 998 RFI candidates and these samples will be utilized to construct the classification model for pulsar candidate selection. At the same time, the performance of the model will be evaluated on the test set which contains 326 pulsar samples and 13321 RFI samples.

\begin{figure*}
\centering
	\subfloat[The diagnostic plots of a pulsar candidate from the FAST data.]{
	\begin{minipage}[b]{0.5\linewidth}
    \centering
    \includegraphics[width=0.4\linewidth]{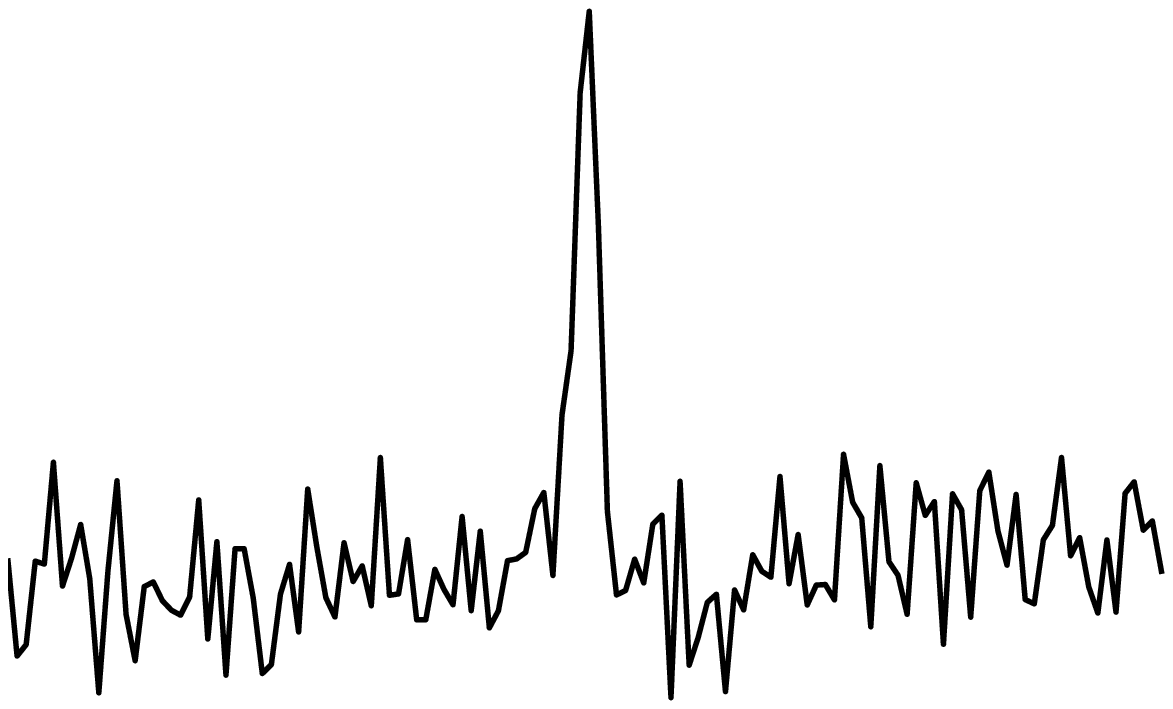}
    \includegraphics[width=0.4\linewidth]{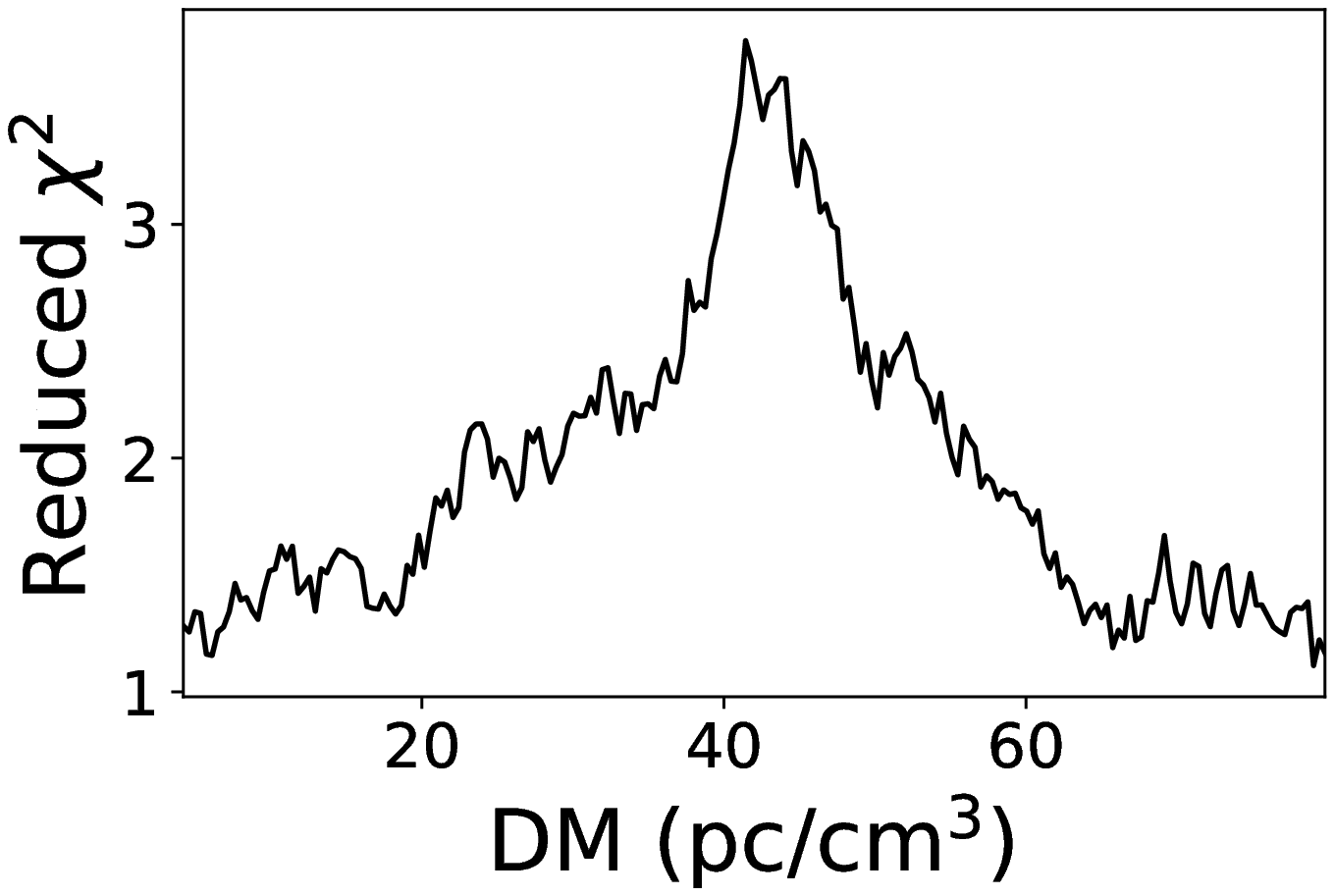}
    \vfill

    \includegraphics[width=0.4\linewidth]{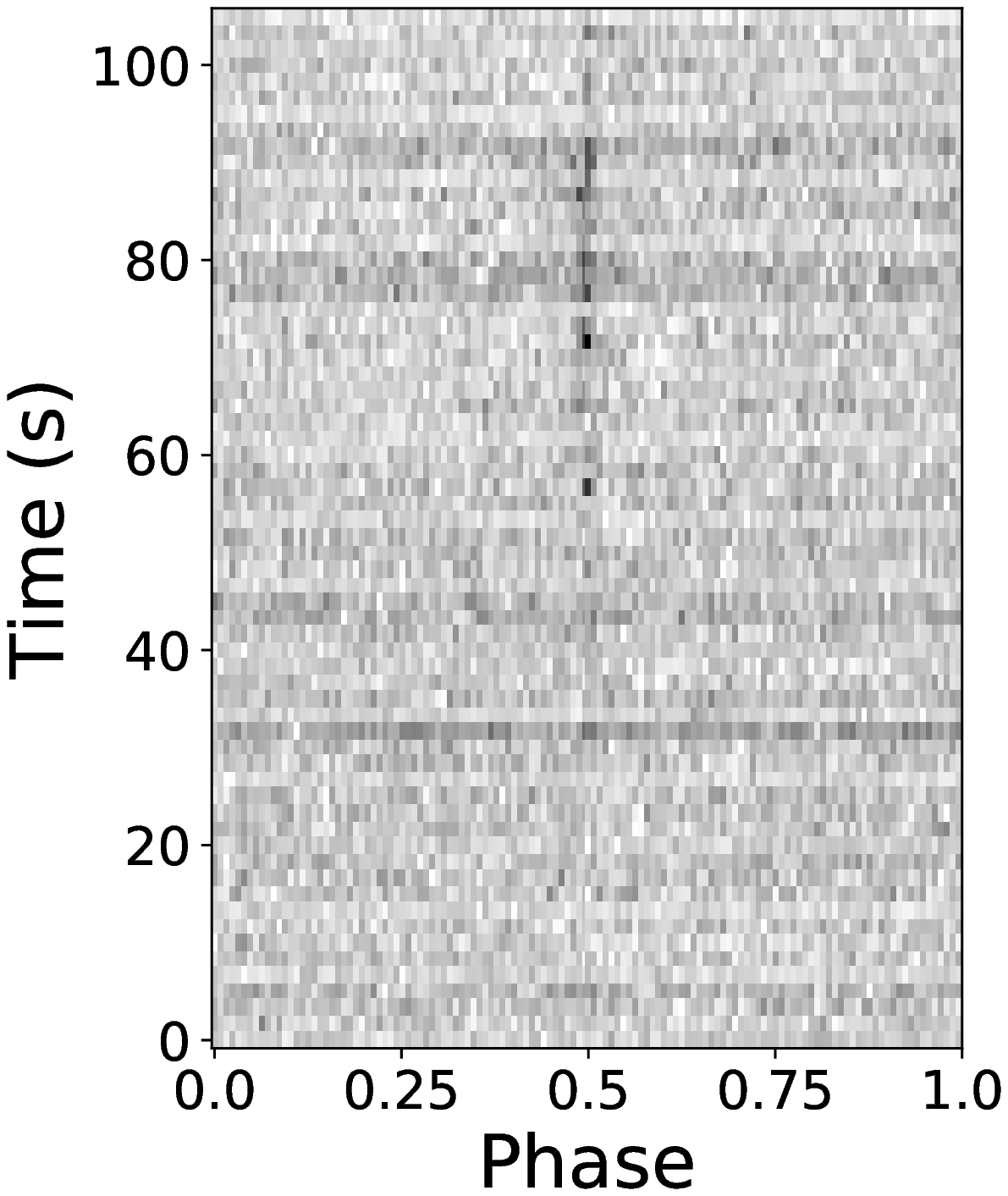}
    \includegraphics[width=0.4\linewidth]{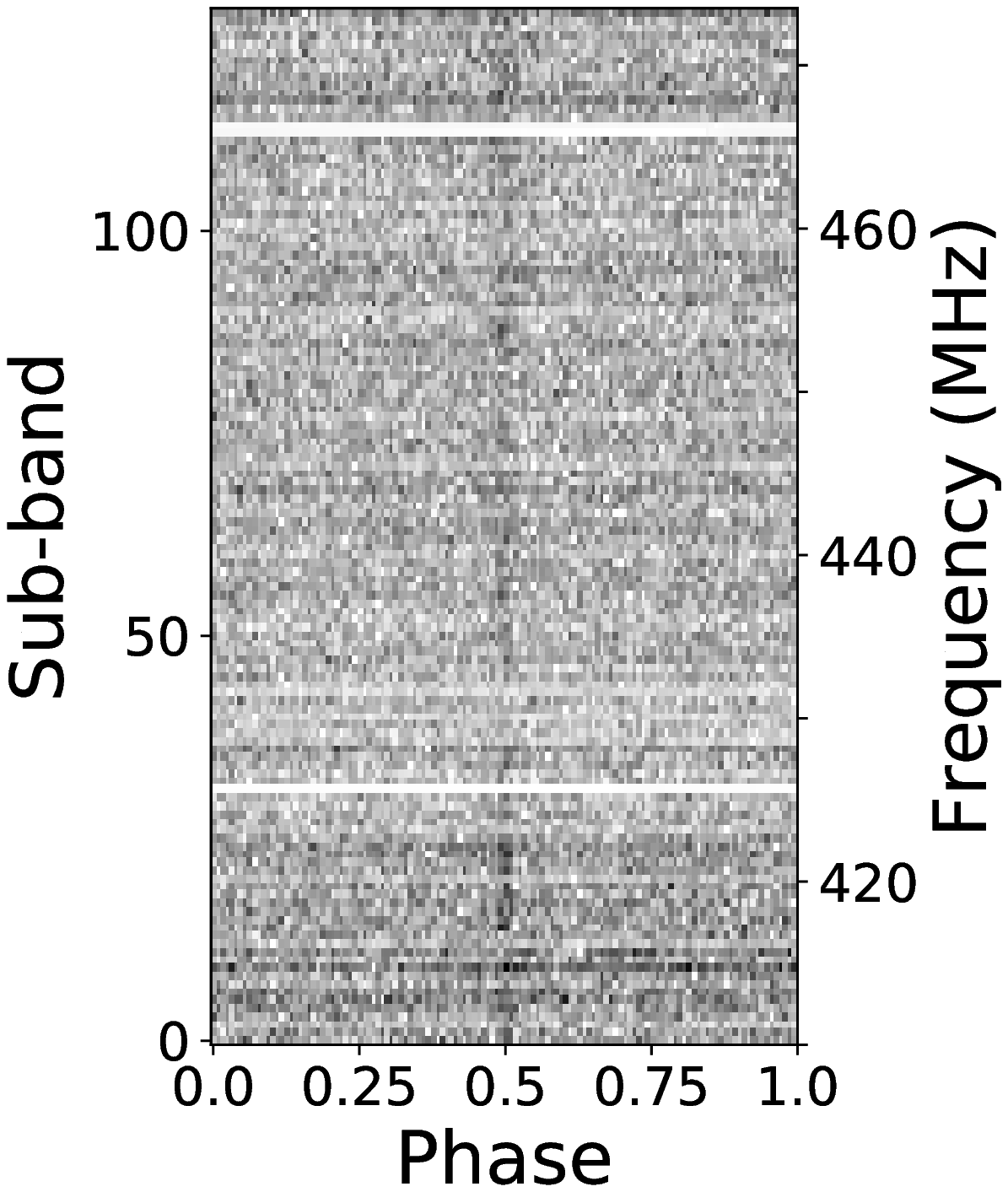}
    \end{minipage}
    \label{fig:pulsar}
    }
    \subfloat[The diagnostic plots of a non-pulsar candidate from the FAST data.]{
	\begin{minipage}[b]{0.5\linewidth}
    \centering
    \includegraphics[width=0.4\linewidth]{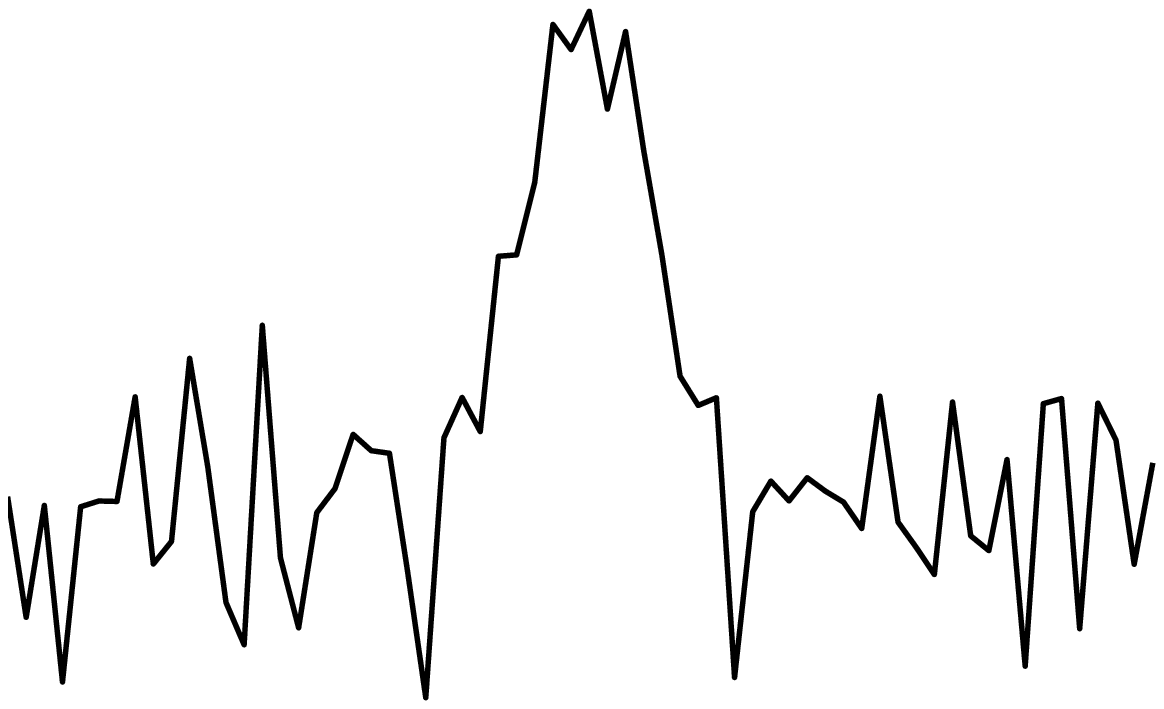}
    \includegraphics[width=0.4\linewidth]{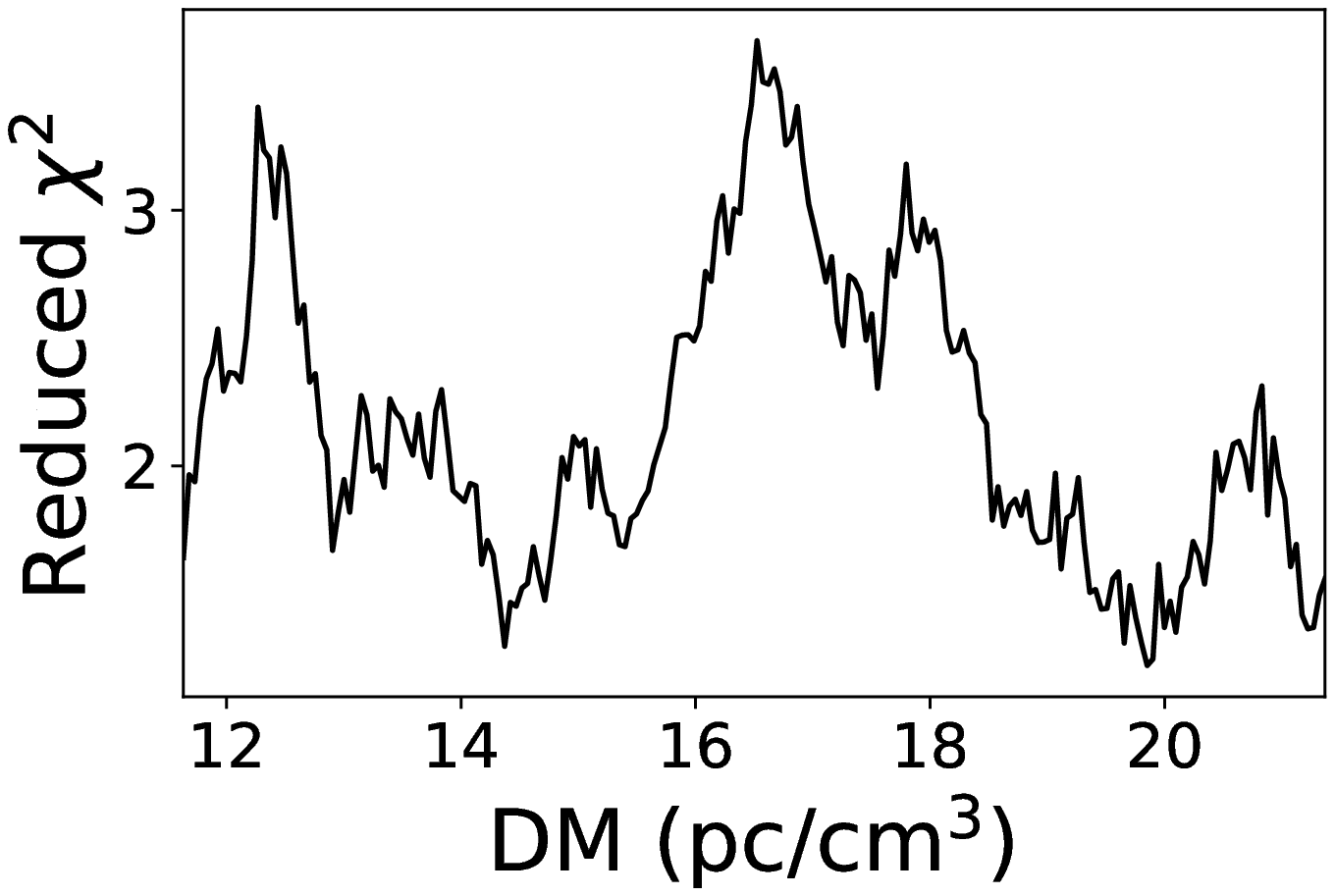}
    \vfill

    \includegraphics[width=0.4\linewidth]{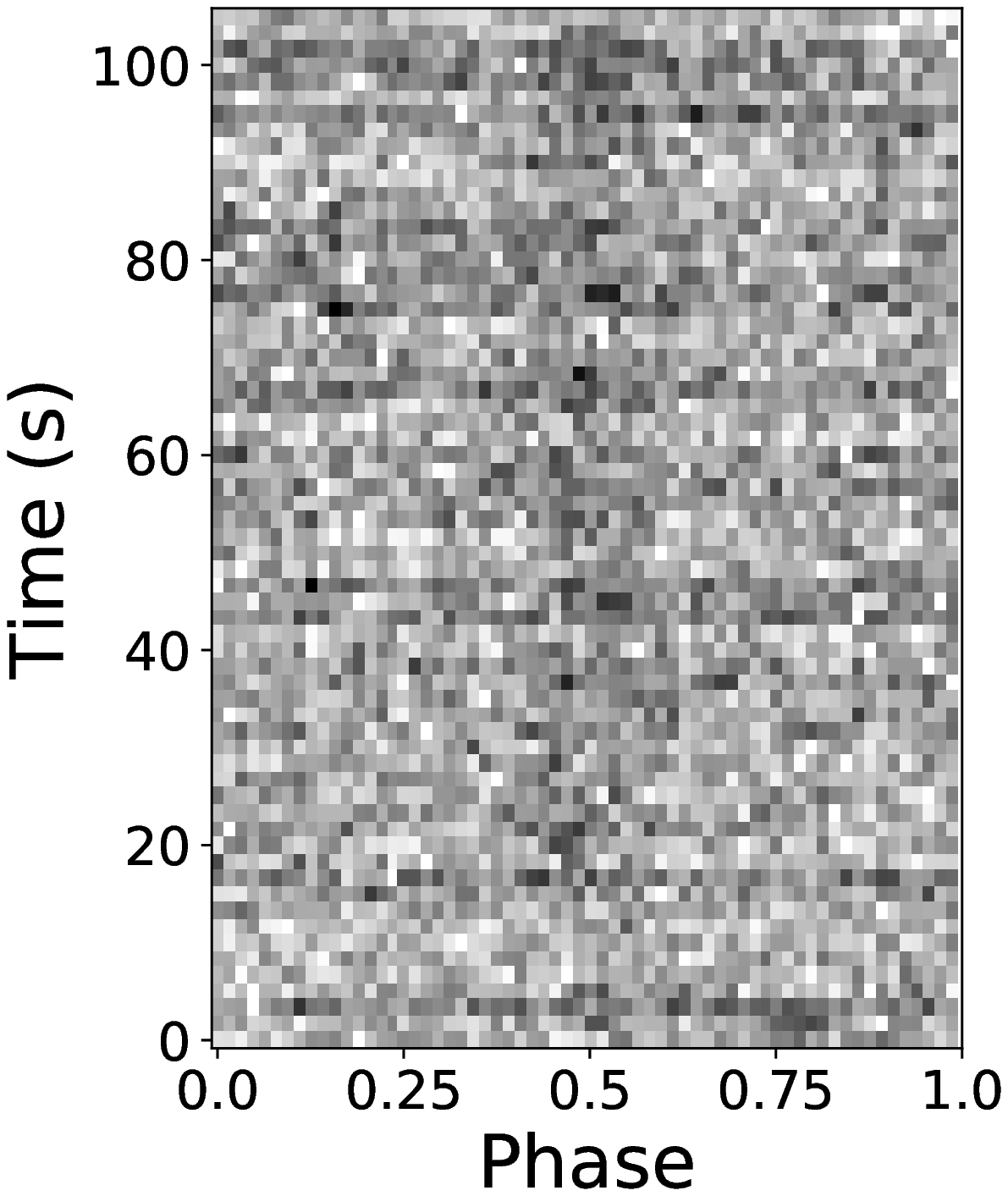}
    \includegraphics[width=0.4\linewidth]{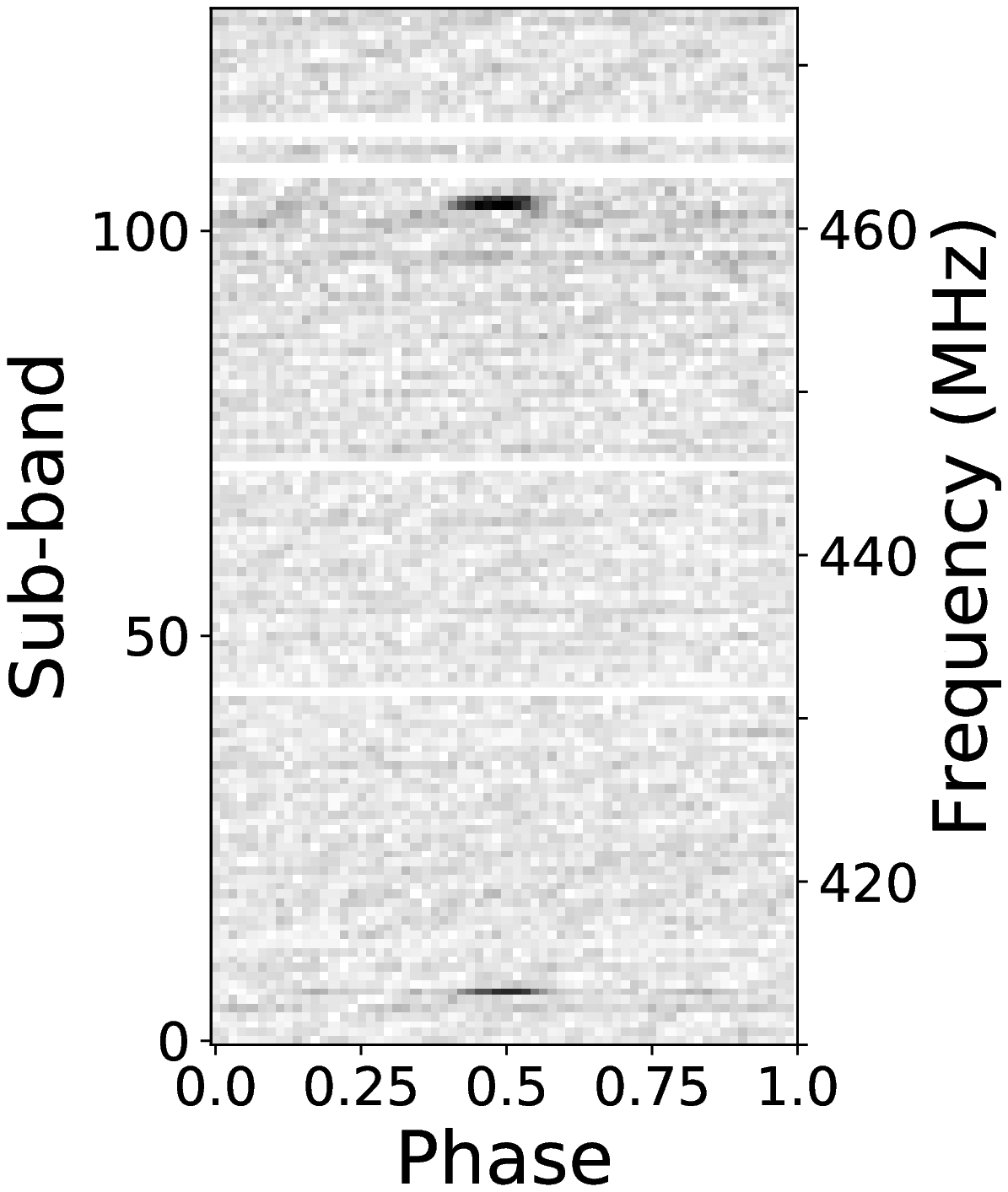}
    \end{minipage}
    \label{fig:rfi}
    }
	
    \caption{Two examples of the pulsar and non-pulsar candidates. For the pulsar candidate, there is a narrow peak in the pulse profile plot and a persistent vertical line both in the time versus phase plot and the frequency versus phase plot. And meanwhile, DM curve peaks at a nonzero value. For the non-pulsar candidate, there are a broad peak in the pulse profile plot and what's more, the pulse only appears in several frequency channels, which indicates that this signal is the RFI.}
    \label{fig:sample}
\end{figure*}

Each sample is processed by PulsaR Exploration and Search Toolkit \citep[PRESTO;][]{ransom2001new, ransom2002fourier} which is a typical software for pulsar search and analysis. After that, the dedispersed and folded three-dimensional (time interval, phase, channel frequency) data is stored in a pfd format file as well as some data descriptions. Summing the data along the frequency channels and time intervals generates the time versus phase plot and frequency versus phase plot. Meanwhile, summing the data along both the time intervals and frequency channels generates the pulse profile histogram. In addition, the last diagnostic subplot is the DM curve which is a plot of the DM trials against the corresponding reduced $\chi^2$ values. Fig.~\ref{fig:sample} presents the diagnostic subplots of a pulsar candidate and a non-pulsar candidate, respectively. For a real pulsar, there should be usually one or more vertical lines in the time versus phase plot and frequency versus phase plot, which indicates a broadband and pulsed signal lasted during the observation time. At the same time, the profile usually contains one or more peaks and the DM curves should peak at a nonzero value. In general, these four diagnostic subplots constitute the fundamental information for the experts to classify the candidates. As a result, they serve as inputs into our model.

\subsection{Data preprocessing}

Before feeding the diagnostic subplots into the model, we have to process the data since there is inconsistency among the candidates, such as the size, scale and so on. These inconsistent factors are useless for identifying the pulsar candidates. What's worse, they have the potential to make negative effects on the training process and performance of the model. Therefore, it is necessary to eliminate these factors before training the model.

Considering the phase-related bias resulting from the peak far away from the center of the plot \citep{zhu2014searching}, we shift the strongest peak to the center phase within the subplots except for the DM curve since the position of the peak is an important pattern for pulsar candidate selection. As a result, the model can pay more attention to the presence of the patterns regardless of their position which is not the necessary factor for identification.

Four diagnostic subplots are all saved as one-dimensional or two-dimensional data arrays, but the size of these arrays vary from candidate to candidate for a certain type of subplots. For the majority of ML algorithms, the size of the inputs should be fixed. Therefore, we have to resize the data arrays to a uniform size: 64 for the pulse profile, 64$\times$64 for the time versus phase plot and frequency versus phase plot and 200 for the DM curve. The plots whose size is smaller than the uniform size are interpolated and those with larger size are scrunched instead of being down-sampled to avoid losing important information. In addition, we normalize the data so that they range from 0 to 1 by using Min-max normalization. The normalization can accelerate the convergence of the gradient descent during training \citep{ioffe2015batch}. On the other hand, normalization does not do any harm to the performance of the model in theory because we just want to extract some certain patterns (e.g. peaks, stripes and so on) from the plots regardless of the exact values in curves or images.

%

\begin{figure}
\centering
	\subfloat[The original DM curve]{
    \includegraphics[width=0.7\linewidth]{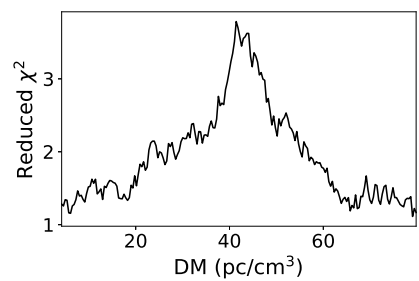}
    \label{fig:HCCNN}}
    \vspace{3ex}

    \subfloat[The generating DM curve]{
    \includegraphics[width=0.7\linewidth]{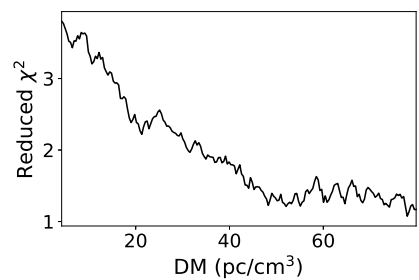}
    \label{fig:VCCNN}}
    \caption{An example for generating a new false positive sample by modifying DM curve coming from a pulsar candidate while other subplots remains unchanged.}
    \label{fig:gen}
\end{figure}

In order to focus the attention of the concatenate layer on the difference between the real pulsar and non-pulsar candidates, we generate some negative samples by replacing only one of the subplots of the pulsar candidates from the training dataset with a corresponding ``non-pulsar subplot''. For example, we firstly choose a pulsar candidate from the training dataset randomly. Secondly, the DM curve is modified by removing the part of the curve before the peak and interpolate the rest part of the curve to the uniform size without modifying the other diagnostic subplots (e.g. Fig.~\ref{fig:gen}). In this way, the generating DM curve peaks at the zero and the newly generated sample belongs to the non-pulsar category. In practice, we can generate the new non-pulsar candidate by modifying any one of the diagnostic subplots of the pulsar candidates except the pulse profile since it can be obtained by summing the frequency versus phase plot over the time intervals.

\begin{figure}
\centering
    \includegraphics[width=0.35\linewidth]{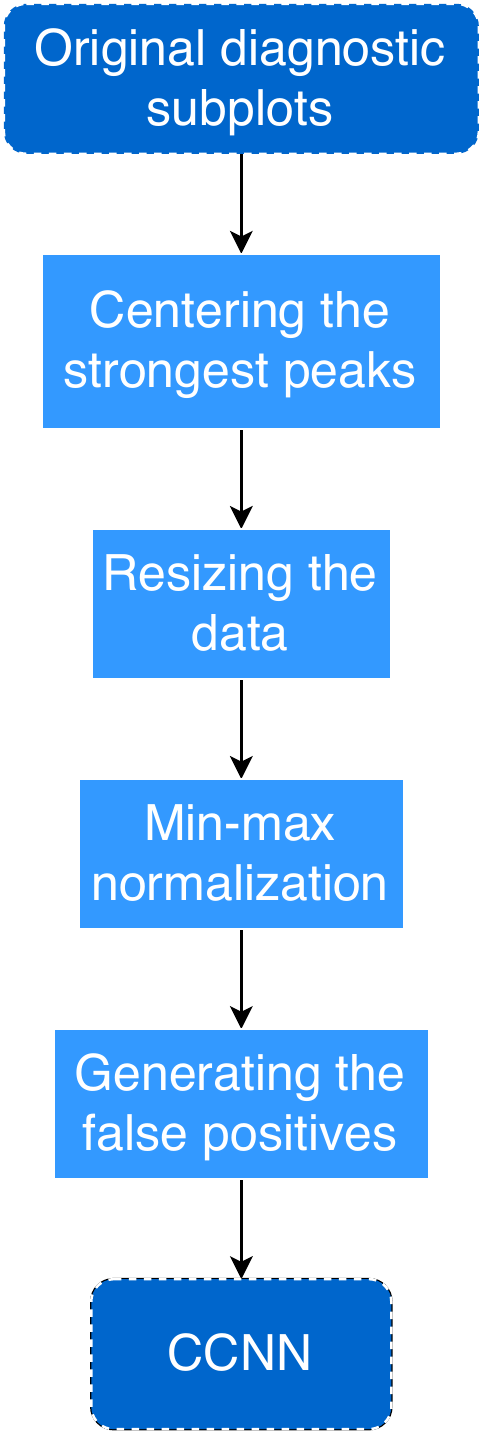}
    \caption{The flowchart of the preprocessing.}
    \label{fig:pre}
\end{figure}

In summary, the flowchart of the data preprocessing is shown in Fig.~\ref{fig:pre}. After that, the processed diagnostic subplots are served as the inputs into the proposed CCNN.


\section{Model}
\label{sec:method}

In this section, the related fundamental components of CCNN are firstly reviewed, including fully connected layer, convolutional layer, global pooling layer; and then the whole architecture of the model is introduced in detail.

\subsection{Fully Connected Layer}

\begin{figure}
\centering
    \includegraphics[width=0.6\linewidth]{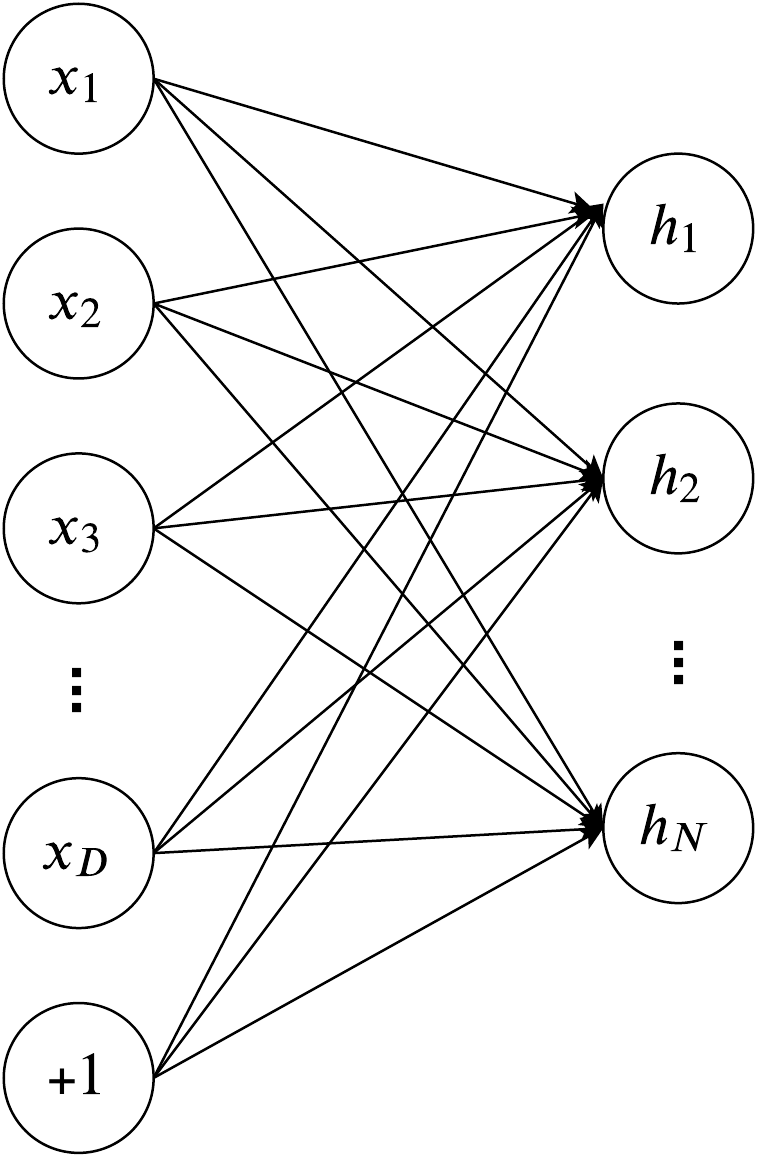}
    \caption{An example architecture of the fully connected layer. The left layer represents the input layer of it and the right one denotes the output layer. The lines connecting these two layers with the arrows pointing from left to the right indicates the weights.}
    \label{fig:fc}
\end{figure}

Fully connected layer is a basic component of the Artificial Neural Network (ANN) which is inspired by the biological neural networks, the fundamental element of animal brains \citep{chen2019design}. An example architecture of this type of layer is shown in Fig.~\ref{fig:fc}.

Generally, a fully connected layer can be considered to be a function composed of some simple mathematical operations which receive an array $\bm{x}\in R^D$ as the input and output another array $\bm{h}\in \mathcal{R}^N$. In detail, all the neurons in the input layer are multiplied by weights (also called synapses) and summed together and then transformed via an activation function\citep{haykin1994neural}. Then, the $i$th output can be expressed as
\begin{equation}
    h_i = \sigma(\bm{w}^\top \bm{x} + b_i)
\end{equation}
where $\bm{w}\in \mathcal{R}^d$ is a weight vector that shows the importance of each input neuron, $b_i\in \mathcal{R}$ is a bias value which allows to shift the function up or down and the $\sigma(\cdot)$ is a activation function which decides whether a neuron should be activated or not. Its motivation is to introduce nonlinearity into the output of a neuron. And then the neural network is able to learn and represent the complex relationship between input data and output target. The activation is triggered by a high similarity between the input data and the patterns stored in weights. Actually, the patterns are unknown or are difficult to be exactly described by humans. Therefore, the weights are initialized randomly. In order to find the correct patterns between the input data and their expected output, it is necessary to adjust the weights. The process of weights adjustment is referred to as learning \citep{yegnanarayana1994artificial}. The goal of learning is to minimize the difference between the neural network outputs and their expected labels. Usually, this process is implemented by back-propagation (BP) algorithm \citep{rumelhart1986learning}.

The structure of the model (e.g. the number of the neurons of fully connected layer and the type of activation function) is chosen by the user and is built by using a simplest one to precisely describe the relationship between the inputs and targets in order to avoid over-fitting \citep{sarle1996stopped}.

\subsection{Convolutional Layer}

\citet{zhu2014searching} creatively introduced two-dimensional CNN to the pulsar candidate selection. The design of CNN is inspired by the work of Hubel and Wiesel who discovered that the cats' visual cortexes contain neurons that individually respond to edges and bars of particular orientations within a small region of the visual field \citep{hubel1959receptive}. The ability of neurons to recognize patterns is unaffected by position shifts.

The working mechanism of convolutional layer is like that one uses a flashlight to slide over a big image from left to right, from top to down. Technically, this flashlight is referred to as filters which actually is a collection of weight vectors, and the area shot by the flashlight is called the receptive field. The output from a neuron is obtained by multiplying the elements of the filter with the image pixels the values in the filter with the original pixel values of the image within the corresponding receptive field and adding all these multiplications together. The activation of the output neuron is triggered if the particular pattern (e.g. the peak and the stripes in diagnostic subplots) are detected from the corresponding receptive field. When the sliding is over, the feature maps (or called activation maps) are obtained. The different filters are used to detect different and simple patterns. As the network going deeper, the patterns extracted by CNN become more complex.

\begin{figure}
\centering
    \includegraphics[width=0.9\linewidth]{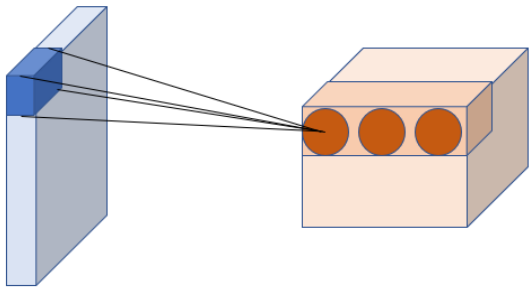}
    \label{fig:cnn}
    \caption{An example architecture of the two-dimensional convolutional layer. The light blue cube represents the input feature map, the blue small cube denotes the filter and the light orange cube represents the output feature maps. Each neuron in the output layer is obtained by multiplying the corresponding elements of the input feature maps and filter.}
\end{figure}

In this work, we not only use the 2D-CNN to extract patterns on the time versus phase plot and frequency versus phase plot, but also process the pulse profile and DM curve using 1D-CNN. The 1D-CNN has been widely applied in time series data, one-dimensional astronomical signals, etc. \citep{pearson2018searching} shows the prominent ability of the 1D-CNN, \citet{zhu2014searching} demonstrates the outstanding performance of 2D-CNN for the time versus phase plot and frequency versus phase plot in pulsar candidate selection. Therefore, the application of 1D-CNN for the other subplots instead of the traditional ML models has the potential of improving the performance of pulsar candidate selection theoretically.

\subsection{Global Pooling Layer}

There is usually a pooling layer in the back of a convolutional layer. The intuitive reasoning behind the pooling layer is that the emphasis of the CNN is to detect the existence of some specific patterns within the image regardless of their exact positions for a classification task. The application of the pooling layer contributes to reducing the number of the parameters thereby decreasing the computational costs. On the other hand, it can effectively alleviate the over-fitting problem.

Pooling layer down-samples feature maps by summarizing each map. Two common pooling methods are average pooling and max pooling. They summarize the average presence of a feature and the most activated presence of a feature, respectively.

In this work, we apply the global pooling which down samples the entire feature map to a single value instead of down sampling patches of the input feature map as the traditional pooling operation does. On the one hand, the global pooling further reduced the number of the training parameters to improve the calculating speed and mitigate over-fitting. On the other hand, after the traditional pooling layer, there is usually a reshape operation that transforms the multi-dimensional arrays to the one-dimensional vectors before being input into the fully-connected layer. This reshape operation may destroy the spatial information in the feature maps. On the contrary, the global pooling is more native to the convolution structure and \citet{lin2013network} has demonstrated that the global pooling has a better performance in some classification tasks than the traditional pooling.

\subsection{Concat Convolutional Neural Network}

\begin{figure*}
\centering
	\subfloat[H-CCNN]{
    \includegraphics[width=\linewidth]{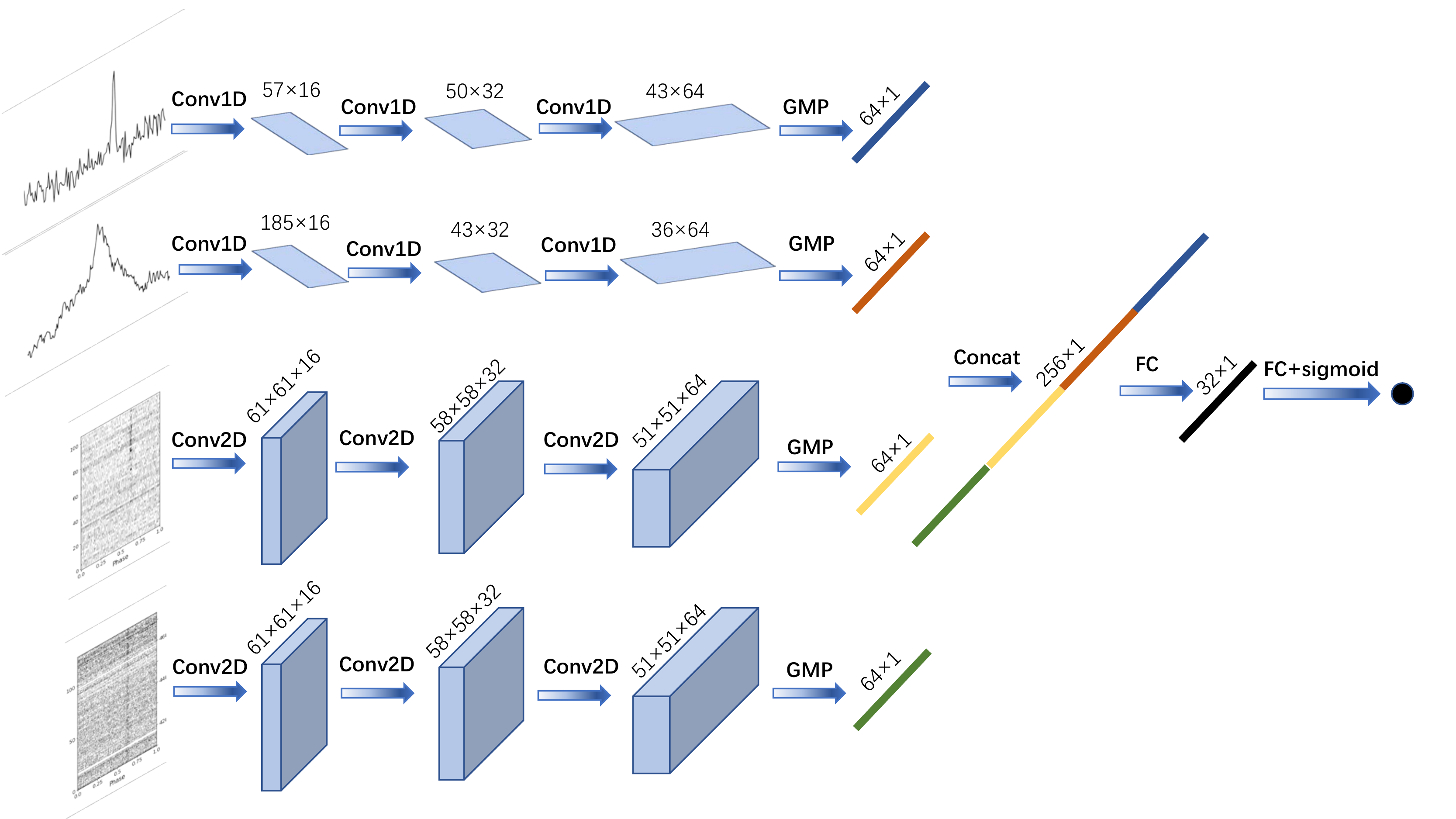}
    \label{fig:H-CCNN}}

    \subfloat[V-CCNN]{
    \includegraphics[width=\linewidth]{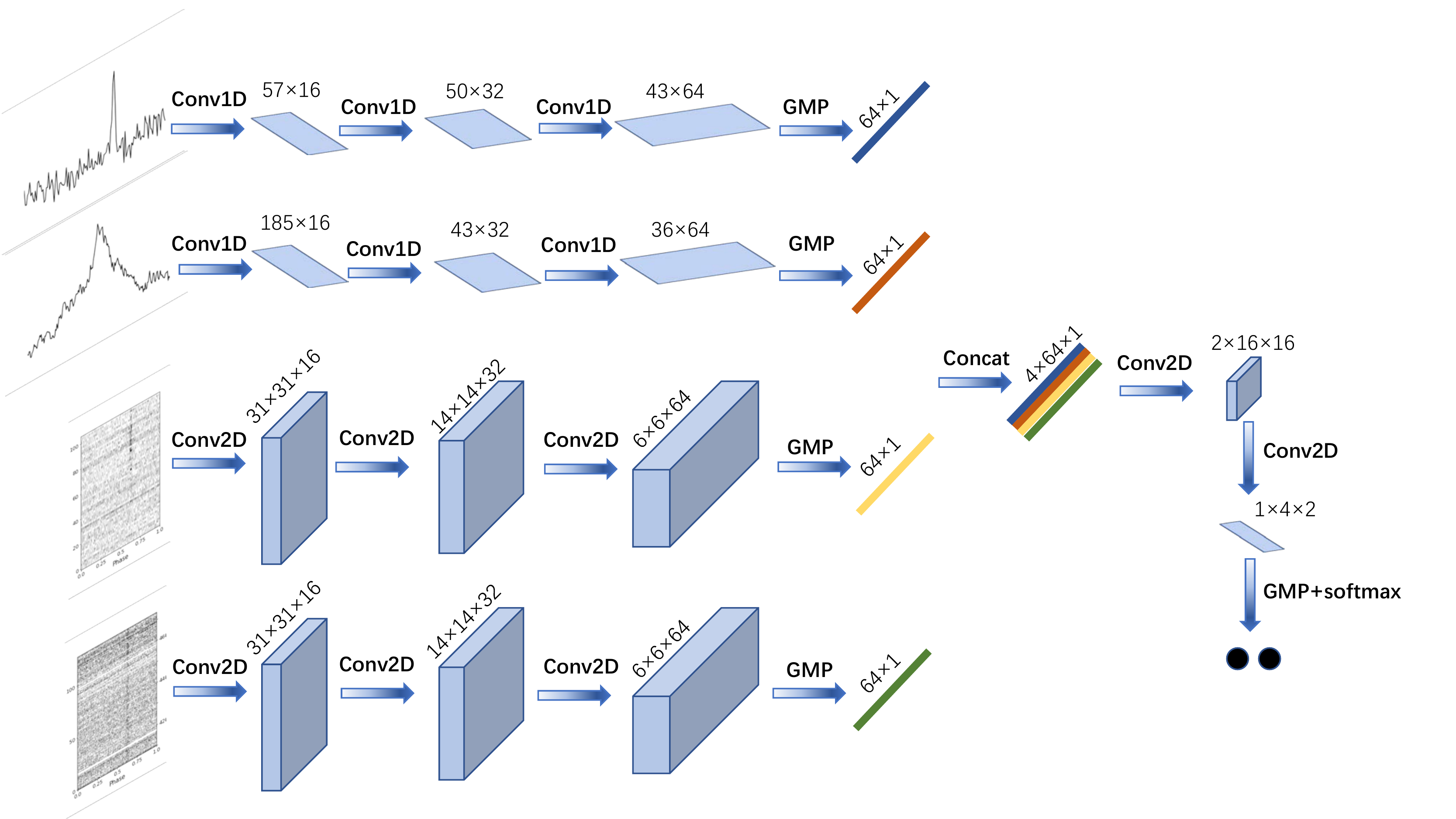}
    \label{fig:V-CCNN}}
    \caption{The architecture of CCNN. The input subplots of CCNN from top to bottom are individually the pulse profile, DM curve, the frequency versus phase plot and the time versus phase plot whose sizes are 64, 200, $64\times 64$, $64\times 64$, respectively. The output sizes of the one-dimensional convolutional layer are $L\times N$ and the output sizes of the two-dimensional convolutional layer are $H\times W\times N$, where $L$ denotes the length of the tensors, $H$ and $W$ are the height and width of the tensors, and $N$ is the number of the feature maps. The output layer of the H-CCNN and V-CCNN are different due to the dissimilar operation in concatenate layer and their subsequent layers. The former model outputs the probability of a candidate being pulsar, while the output layer of the latter model contains two neurons repectively representing the probabilities of a candidate being pulsar and non-pulsar. GMP means Global Max Pooling layer, Concat represents the Concatenate layer and FC is the abbreviation of  Fully connected layer.}
    \label{fig:CCNN}
\end{figure*}

In this work, we use four convolutional neural networks to extract features respectively from four diagnostic subplots: two 1D-CNNs respectively for the pulse profile and DM curve, and two 2D-CNNs respectively for the time versus phase plot and frequency versus phase plot. Considering that the extracted features, such as peaks or the vertical stripes, are some simple pattern, the CNN for each subplot in this work only consists of three convolutional layers and is followed by a global max pooling layer to summarize the information of the feature maps and output a one-dimensional vector. And then, a concatenate layer is applied to merge the information coming from four different diagnostic subplots.
The proposed scheme is referred to as Concat Convolutional Neural Network (CCNN), and this work used two examples of the CCNN: Horizontal Concat Convolutional Neural Network (H-CCNN) (Fig. \ref{fig:H-CCNN}) and Vertical Concat Convolutional Neural Network (V-CCNN) (Fig. \ref{fig:V-CCNN}).

The difference between H-CCNN and V-CCNN are on their concat type and subsequent layers. In detail, the former model concatenates the four vectors extracted from the four diagnostic subplots one after another in horizontal direction to form a long one-dimensional vector; while the latter concatenates them in vertical direction to generate a two-dimensional matrix. After that, the final two layers of H-CCNN are the fully connected layer and the activation of the last one is sigmoid function. The sigmoid activation function is to compute the probabiliy of a candidate being pulsar. The subsequent layers of V-CCNN are several convolutional layers and one global average pooling layer which averages each feature maps; and the resulting vector of the global average pooling layer is fed into a softmax layer \citep{lin2013network}. Two neurons in the output layer stand for the probabilities that the candidate is of pulsar and non-pulsar, respectively. The choice of activation function for the last layer and the parameters for the convolutional layers of H-CCNN and V-CCNN were determined by their $F_1$ score performance. This work investigated sigmoid function and softmax function for finding the appropriate activation. And the output of sigmoid function is a real value, denoted by $p$ for convenience, between 0 and 1. This value represents the probability of a candidate being pulsar. Therefore, the probability of the candidate being a non-pulsar can be calculated by $1-p$. A brief architectures of CCNN (H-CCNN and V-CCNN) are presented in Fig.~\ref{fig:CCNN}.

The configuration of CCNN (e.g. the number and size of the filters, the type of the global pooling, the number of the neurons in the fully connected layer, the type of the activation function, etc.) is determined by grid search based on 10-fold cross-validation \citep{james2013introduction}. Firstly, the training dataset is shuffled randomly and split into 10 groups. Each unique group is taken as the validation set one after another and the corresponding remaining data is served as the training set which is used to train the CCNN. The performance is evaluated by computing the $F_1$ score on the validation set. Finally, the hyperparameters are determined based on the model with the highest $F_1$ score. As a result, the optimal structure of CCNN is shown in Fig.~\ref{fig:CCNN}, and the model is trained by an Adam optimizer \citep{kingma2014adam} with learning rate of 0.001 and batch size of 64.

The CCNN is implemented using Keras \citep{chollet2015keras} with the TensorFlow backend \citep{tensorflow2015-whitepaper}\footnote{https://github.com/xrli/CCNN/}. Keras is high-level neural networks API written in Python and it focuses on enabling fast experimentation instead of coding ability. This characteristic of Keras makes it suitable for the process and analysis of the astronomical data.

\section{Experimental Investigation}
\label{sec:Results_Discussion}

To investigate the effectiveness of the CCNN, some quantitative evaluations and comparisons are conducted on FAST observations for pulsar candidate selection. In this section, we first introduce evaluation metrics, and then present and analyse the experimental results.

\subsection{The evaluation metrics}
In pulsar candidate selection, the pulsars and their harmonic signals are served as positive samples, while the remainings, non-pulsars, are regarded as negative samples. Therefore, this problem can be considered as a binary classification task. The common evaluation metrics for the binary classification problem include Accuracy, Precision, Recall and $F_1$ score. All the metrics utilized in this work are defined as follows.

\begin{align}
Accuracy &= \frac{TP+TN}{TP+TN+FP+FN},\\
Precision &= \frac{TP}{TP+FP},\\
Recall &= \frac{TP}{TP+FN},\\
F_1 \ score &= \frac{2\times Precision \times Recall}{Precision + Recall},
\end{align}
where TP, FP, TN and FN respectively denote the number of True Positive (both ground-true label and prediction are pulsars), False Positive (ground-true label is the non-pulsar and prediction is pulsar), True Negative (both ground-true label and prediction are non-pulsar), and False Negative (ground-true label is pulsar and prediction is non-pulsar).

Accuracy indicates the fraction of the correct predictions on the whole. Precision and Recall are inversely proportional respectively to the FP and FN. Therefore, the Precision and Recall indicate the severity of false detection and missed detection, respectively. The false positives will waste labor and time for the further observation while the false negatives make negative effects on the search for the new pulsars. Considering the test set is heavily imbalanced, $F_1$ score is also used to assess the performance of the model \citep{jeni2013facing}. $F_1$ score is defined as the harmonic mean of precision and recall and is served as a trade-off between them. All of the metrics range in $[0, 1]$. The smaller the FPR and FNR, the better the RFI detection scheme. On the contrary, a higher value of accuracy and F1 score are more satisfying.


\subsection{The experiment on FAST data}

The experimental data collected from the FAST drift-scan survey has been split into the training set and test set when it was public, where there are only 1835 candidates in total, among which 837 are pulsars or their harmonics and 998 are non-pulsars.

For a deep learning model, less training data is more likely to make it over-fitting. In statistics, over-fitting refers to an analysis that matches a particular and known data set exactly but fails to fit the other data well or predict future observations satisfying \citep{leinweber2007stupid}. In ML, an over-fitting model usually possesses too many parameters or heavily complex structures for the limited data \citep{tetko1995neural}. As the epochs of training increases, the performance of an over-fitting model keeps improving on the training set but will degrade after a period of growth on the test set. To overcome this problem, we adopt early stopping during the training process. Firstly, the original training set is randomly split into two groups: 80\% for training and 20\% for performance validation. Secondly, the performance of the model is evaluated on the validation set at the end of each epoch. At the same time, the loss at the current epoch is compared to the previous one or that of the saved model, the model with smaller loss will be saved. And then, stopping the training process will be triggered when the loss score of the validation set increases for successively five times.

\begin{figure*}
\centering
	\subfloat[The training process of H-CCNN]{
    \includegraphics[width=0.45\linewidth]{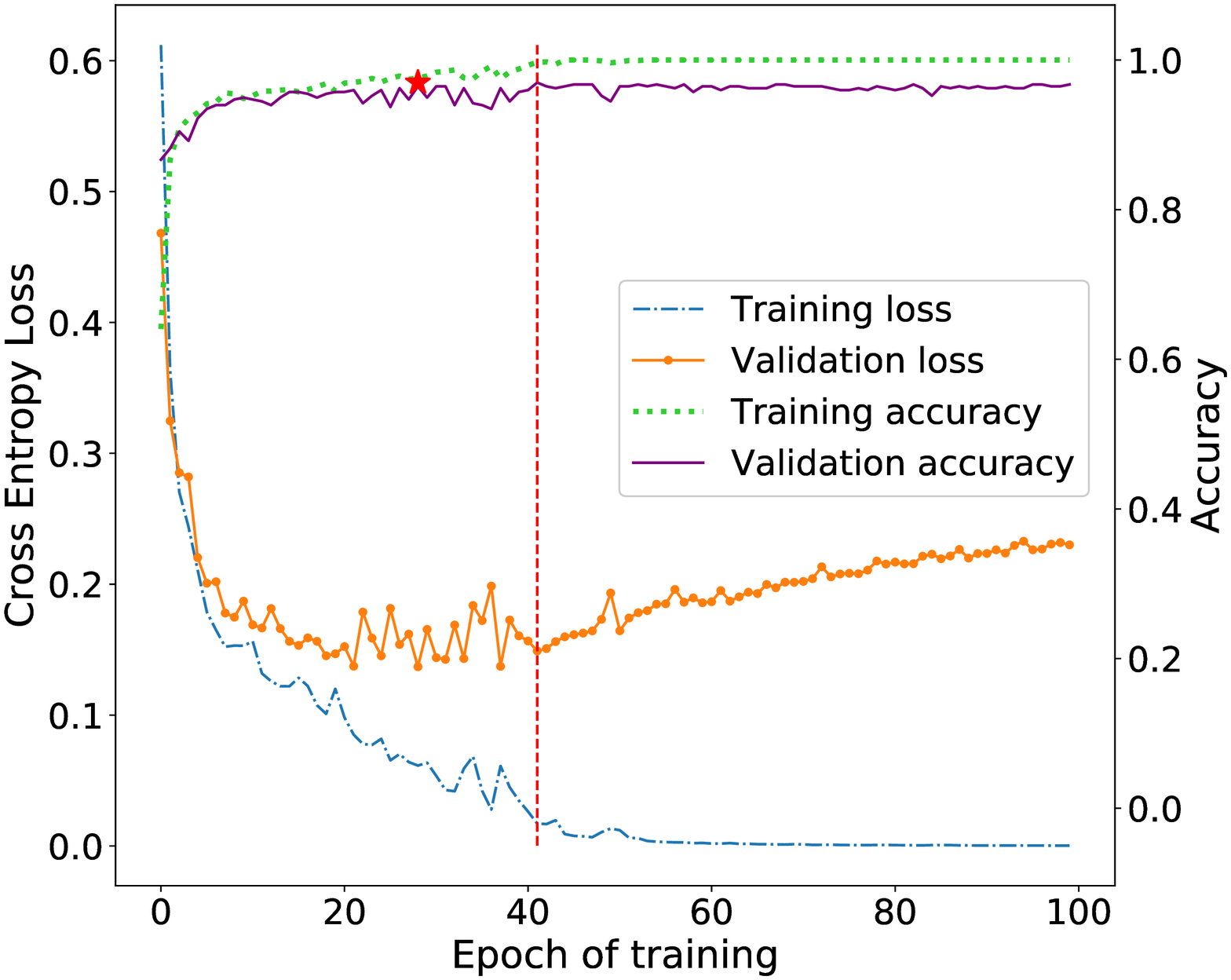}
    \label{fig:HCCNN}}
    \hspace{6ex}
    \subfloat[The training process of V-CCNN]{
    \includegraphics[width=0.45\linewidth]{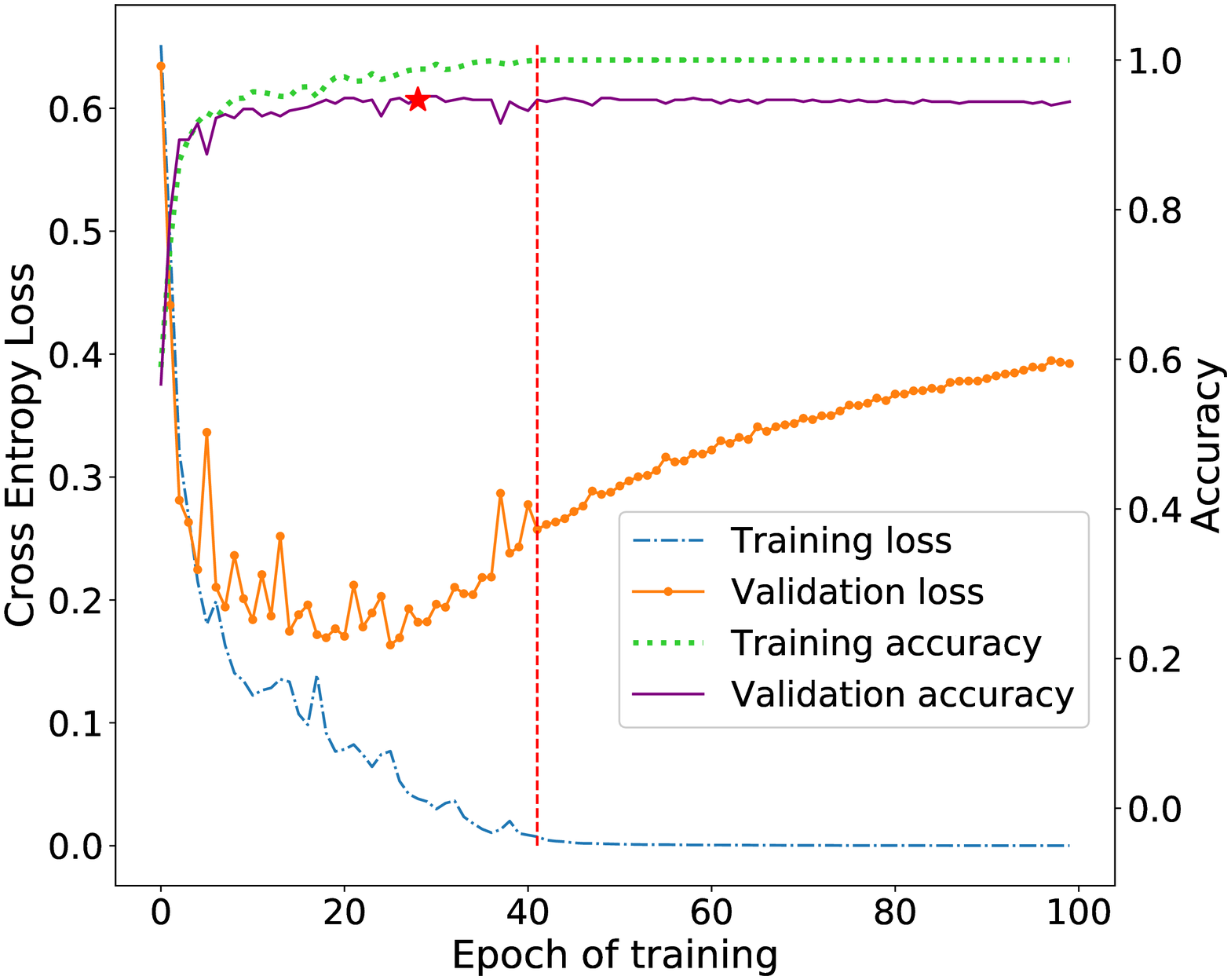}
    \label{fig:VCCNN}}
    \caption{The loss and accuracy curves of training and validation set during the training. The vertical dashed line means the trigger point of early stopping after which the validation loss increases almost continuously. The star-like marker shows the best performance of the model on validation set and locates the point before the trigger point of early stopping.}
    \label{fig:CCNN_training}
\end{figure*}

Fig.~\ref{fig:CCNN_training} shows the trends of the loss and accuracy during the training. It is worth noting that we plot the whole training process with 100 epochs to facilitate the observation and in fact, the training process has been early stopped at the 46th epoch. The trends of the loss and accuracy curves of both the H-CCNN and V-CCNN share similar characteristics. Therefore, we just take that of H-CCNN for a detailed analysis. The changing trends of the loss curves are consistent with the above analysis, which indicates that both H-CCNN and V-CCNN become over-fitting as the training continues after the trigger-point. Therefore, it is necessary to use early stopping during the training process for alleviating the problem of over-fitting. The curves in Fig. \ref{fig:CCNN_training} indicate that the model becomes over-fitting after the 46th epoch. The validation accuracy curve peaks at the 41th epoch and keeps oscillating after that, which means that the model achieves its best performance at this epoch and it is saved as the final model used to make the classification on the test set. And two final models are compared to the state-of-the-art ML models: PICS \citep{zhu2014searching} and PICS-ResNet \citep{wang2019pulsar}, which were trained in the previous works and have been public\footnote{ \url{https://github.com/dzuwhf/PICS-ResNet/tree/master/ubc_AI}}.

The results of these models tested on the FAST data are presented in Table~\ref{tab:results}. On the whole, H-CCNN achieves the best performance on the FAST data, especially on the accuracy, precision, and $F_1$ score. In detail, the recall of the H-CCNN, 0.9635, is a higher than that of PICS and is relatively less than that of PICS-ResNet, 0.9816. However, H-CCNN achieves over 10 percentage points precision higher than PICS as well as PICS-ResNet. It is because that PICS mistakenly identifies 863 non-pulsars as the pulsar signals while the number of the false positives of H-CCNN is approximately half of that. It means that the ability of PICS and H-CCNN to correctly identify the real pulsar signals among all the candidates collected from FAST are almost the same but at the same time, H-CCNN is far better at excluding false candidates than PICS when identifying the pulsars. As a result, using H-CCNN for pulsar candidate selection in the practical application can reduce the labor and expense for the further observations. On the other hand, V-CCNN achieves the highest recall among all the classifiers and only misses 4 pulsars in all. In addition, we average the scores from the final H-CCNN and V-CCNN, then the mean scores are served as the final score of H-CCNN+V-CCNN. This operation is served as ensemble learning in the field of ML. Ensemble learning combines the results coming from multiple different learning algorithms to obtain a better performance than that of any single one component alone \citep{rokach2010ensemble}. In this way, H-CCNN+V-CCNN inherits high precision from H-CCNN and high recall from V-CCNN. Therefore, the results show that all the metrics of H-CCNN+V-CCNN are higher than that of PICS, and are equal to or higher than that of PICS-ResNet. In summary, CCNN outperforms the PICS and the PCIS-ResNet on the FAST data in general. It is worth noting that CCNN is trained only on FAST data while the PICS and PICS-ResNet are trained in three datasets, including PALFA, HTRU and FAST. In theory, the more training samples for a deep neural network, the better the performance of the network. Therefore, it is likely to improve the performance of the CCNN if the other two datasets are available for training.

\begin{table*}
	\centering
	\caption{The evaluation results of four classifiers on the FAST data set. Note that the first two models were trained by \citet{guo2019pulsar}. The final model, ``H-CCNN+V-CCNN'', is the embedding model of H-CCNN and V-CCNN. The Boldface digits indicate the best performance. }
	\label{tab:results}
	\begin{tabular}{llccccc} 
		\hline
		Model &Training dataset (No. training samples) & Accuracy & Precision & Recall & $F_1$ score & No. missing pulsars\\
		\hline
		PICS & FAST + HTRU + FALFA (13632) & 0.9357&0.2649 & 0.9540& 0.4146&15\\
		PICS-ResNet & FAST + HTRU + FALFA (13632)&0.9332 &0.2612 &0.9816 &0.4126&6\\
		H-CCNN & FAST (1835) & $\bm{0.9634}$& $\bm{0.3920}$&0.9632&$\bm{0.5572}$ &12\\
		V-CCNN & FAST (1835)&0.9173&0.2227&$\bm{0.9877}$& 0.3634&$\bm{4}$\\
		H-CCNN+V-CCNN&FAST (1835) & 0.9476 &  0.3110 & 0.9816 &  0.4723&6\\
		\hline
	\end{tabular}
\end{table*}


\subsection{The analysis of the missing pulsars}

\begin{figure*}
\centering
	\subfloat[Missing pulsar 1]{
	\begin{minipage}[b]{0.5\linewidth}
    \centering
    \includegraphics[width=0.4\linewidth]{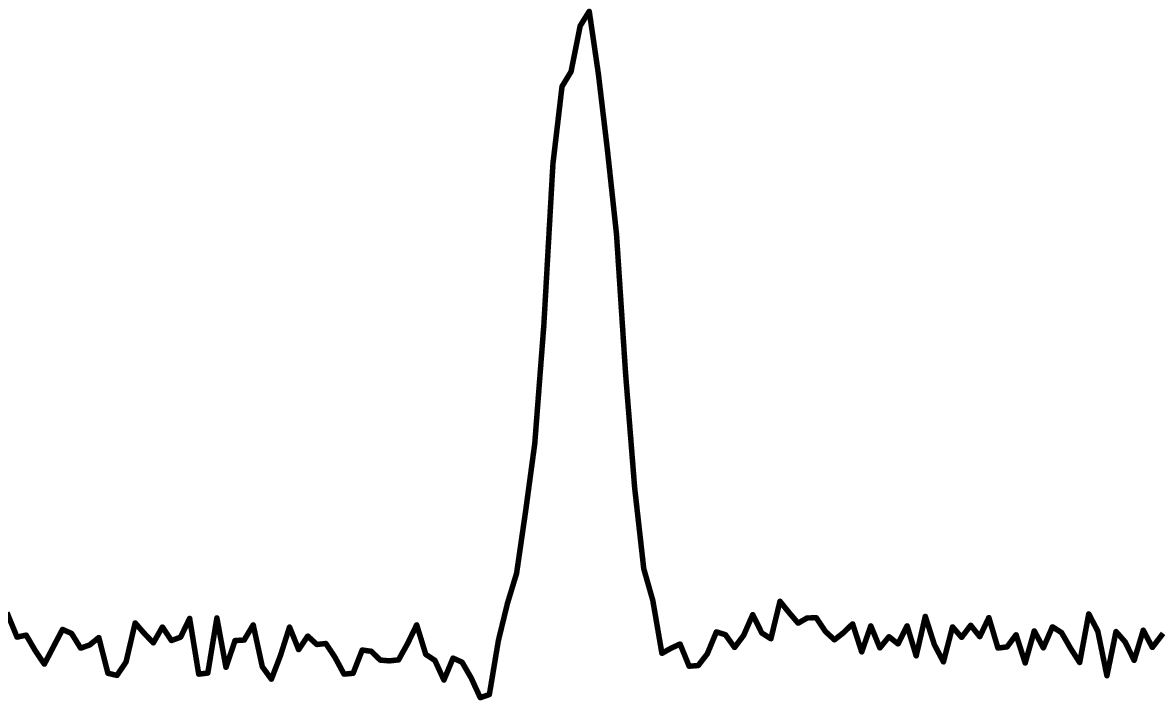}
    \includegraphics[width=0.4\linewidth]{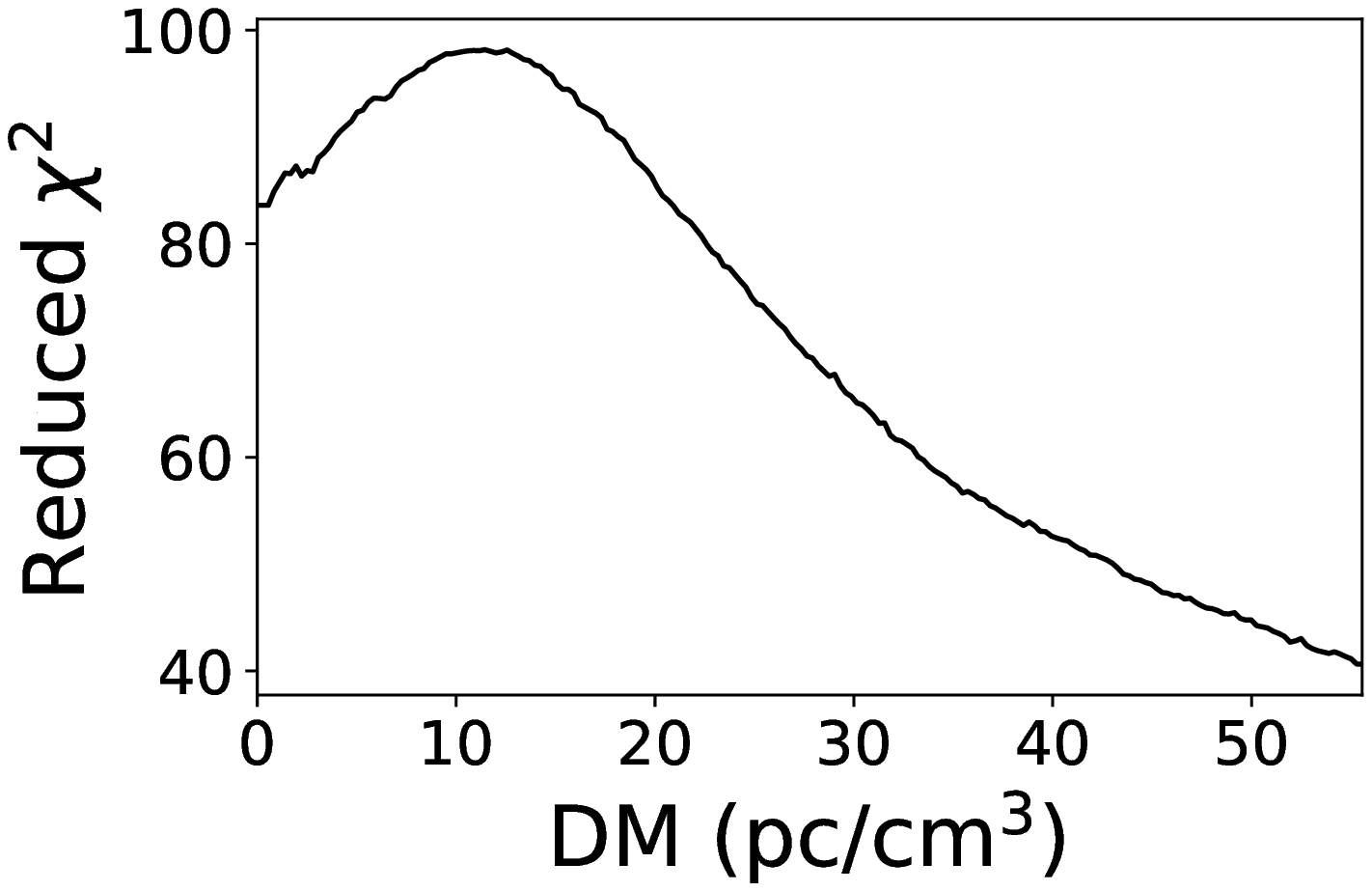}
    \vfill

    \includegraphics[width=0.4\linewidth]{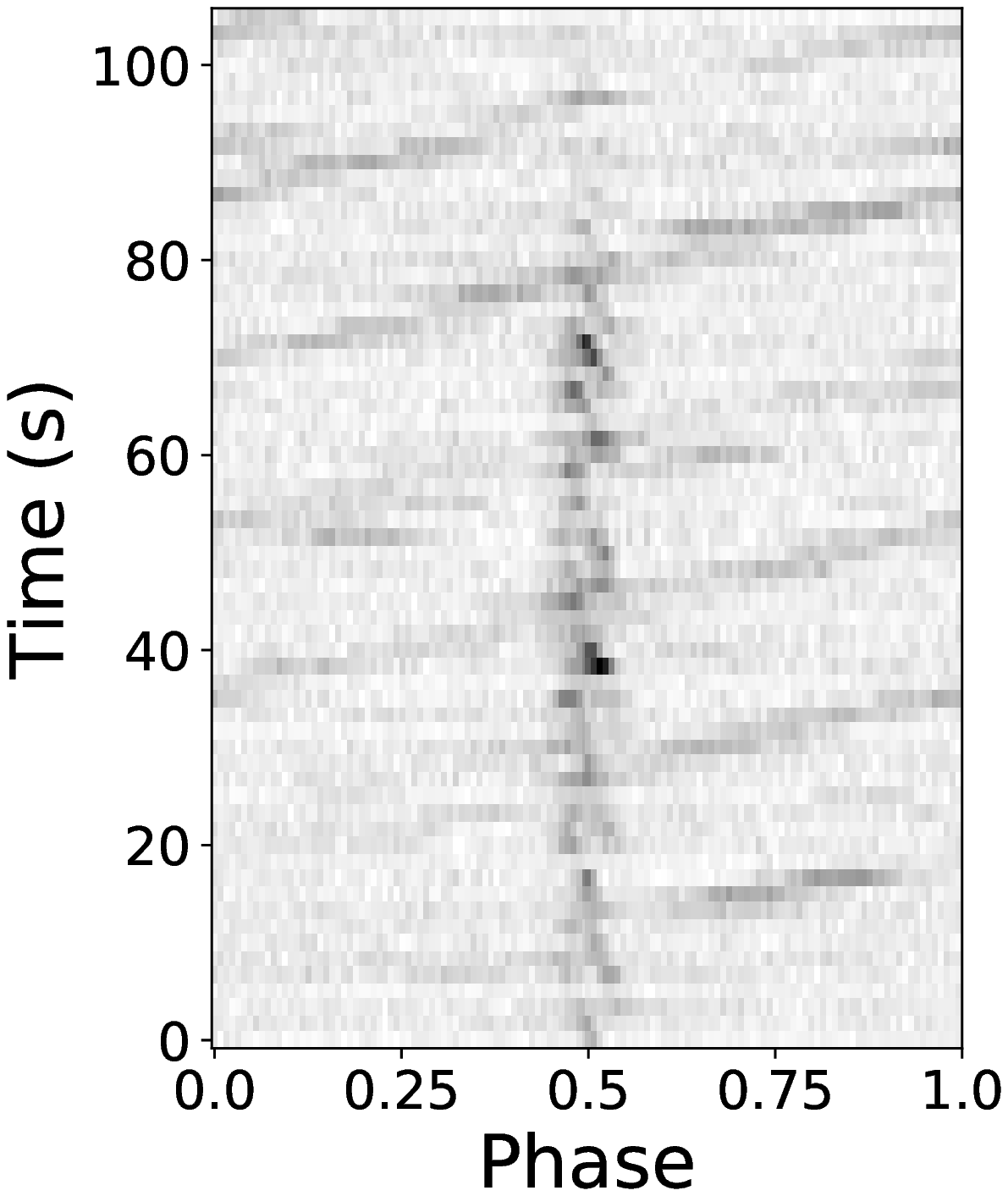}
    \includegraphics[width=0.4\linewidth]{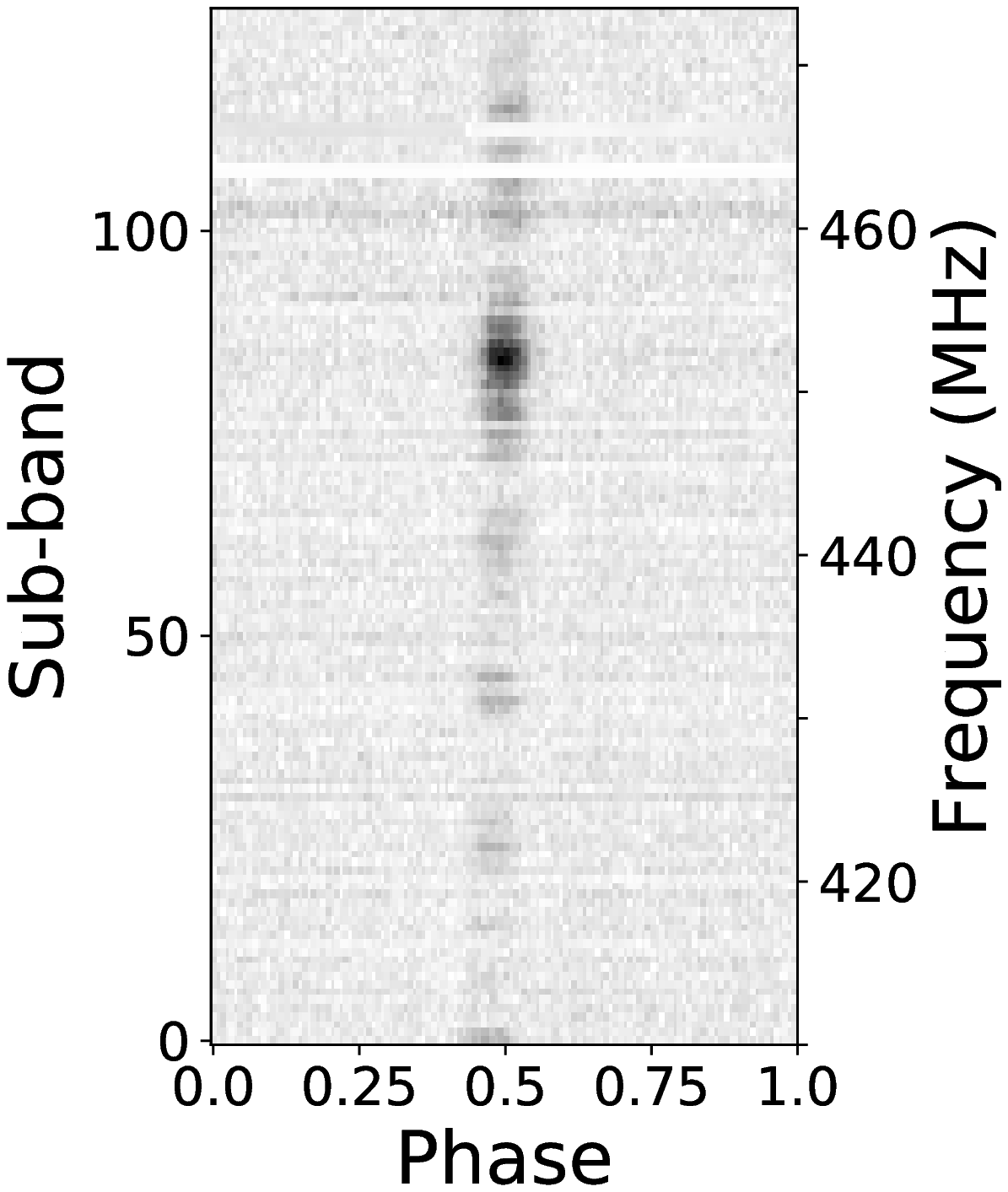}
    \end{minipage}
    \label{fig:wrong1}
    }
    \subfloat[Missing pulsar 2]{
	\begin{minipage}[b]{0.5\linewidth}
    \centering
    \includegraphics[width=0.4\linewidth]{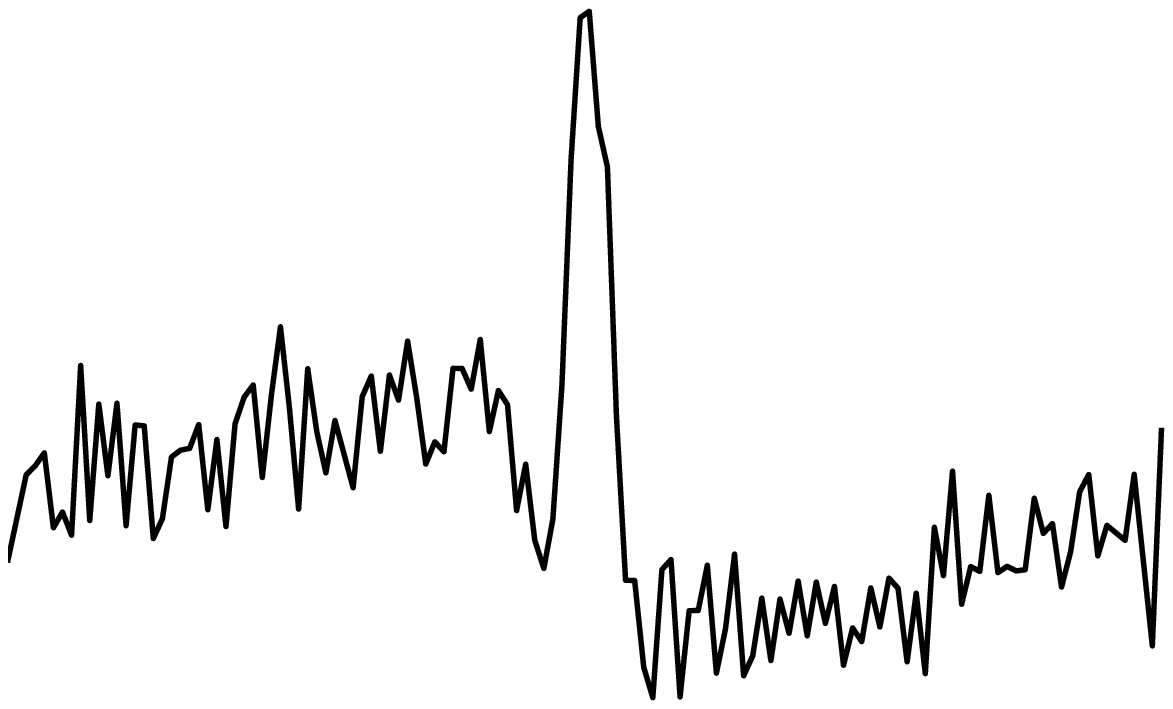}
    \includegraphics[width=0.4\linewidth]{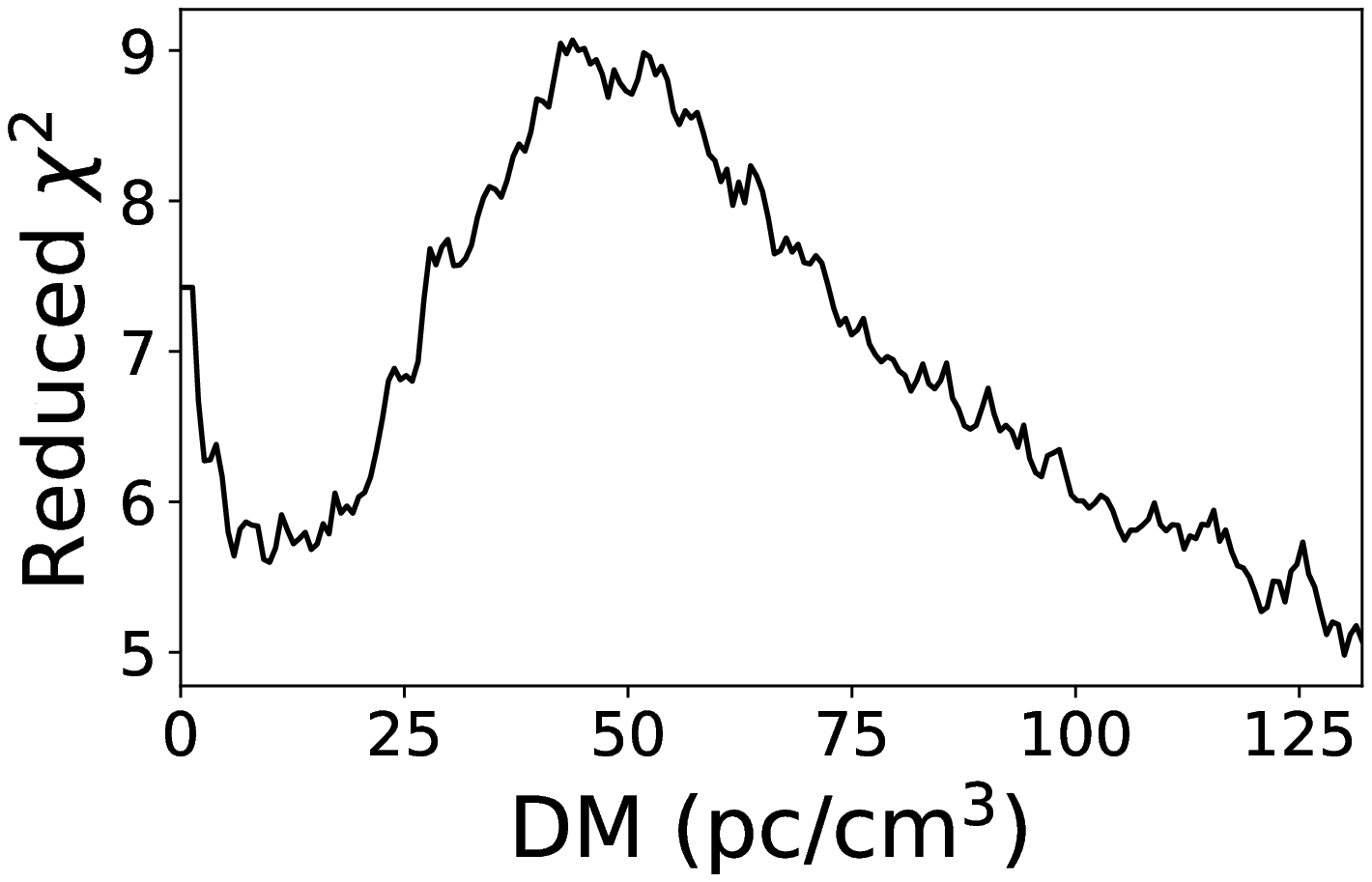}
    \vfill

    \includegraphics[width=0.4\linewidth]{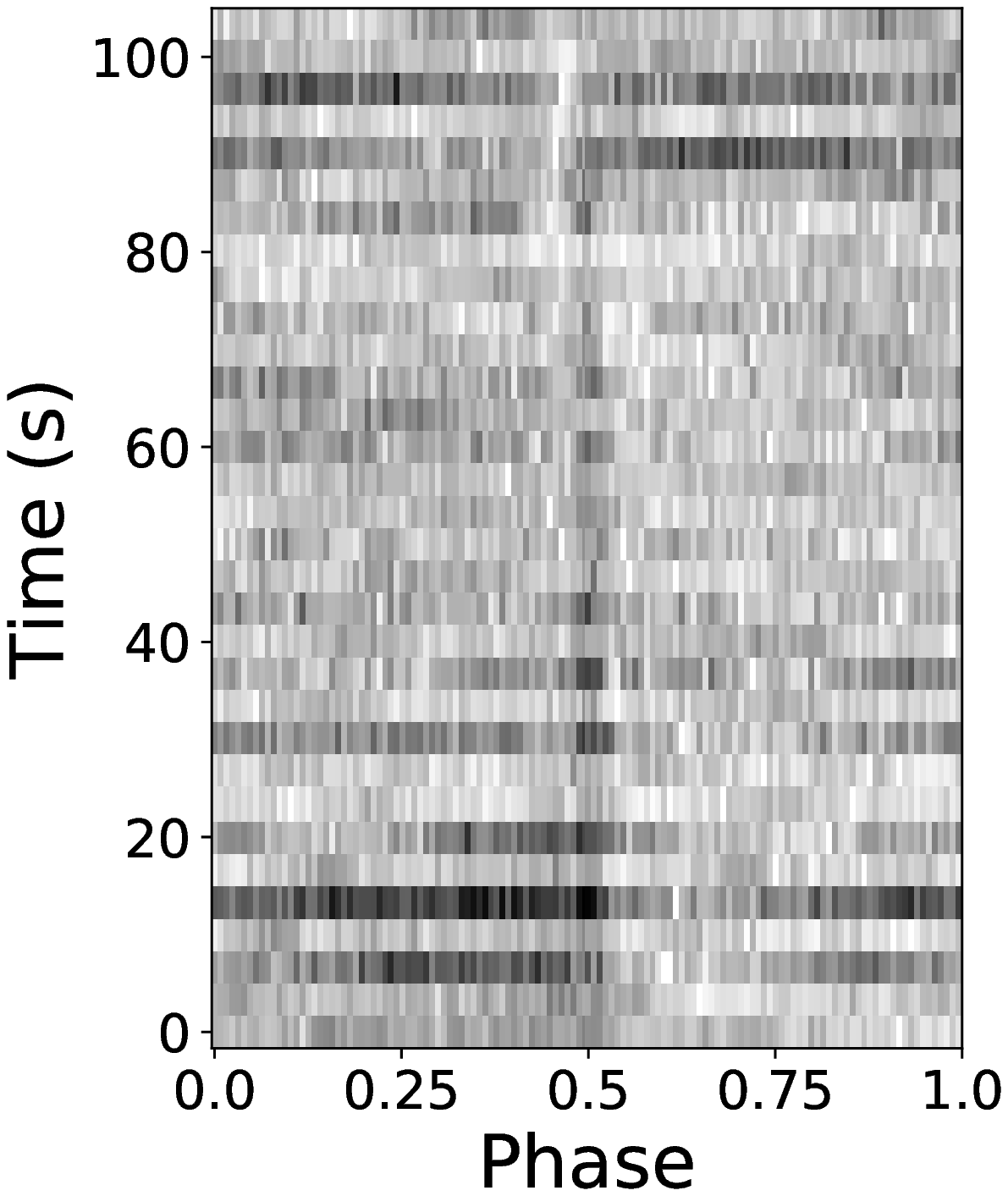}
    \includegraphics[width=0.4\linewidth]{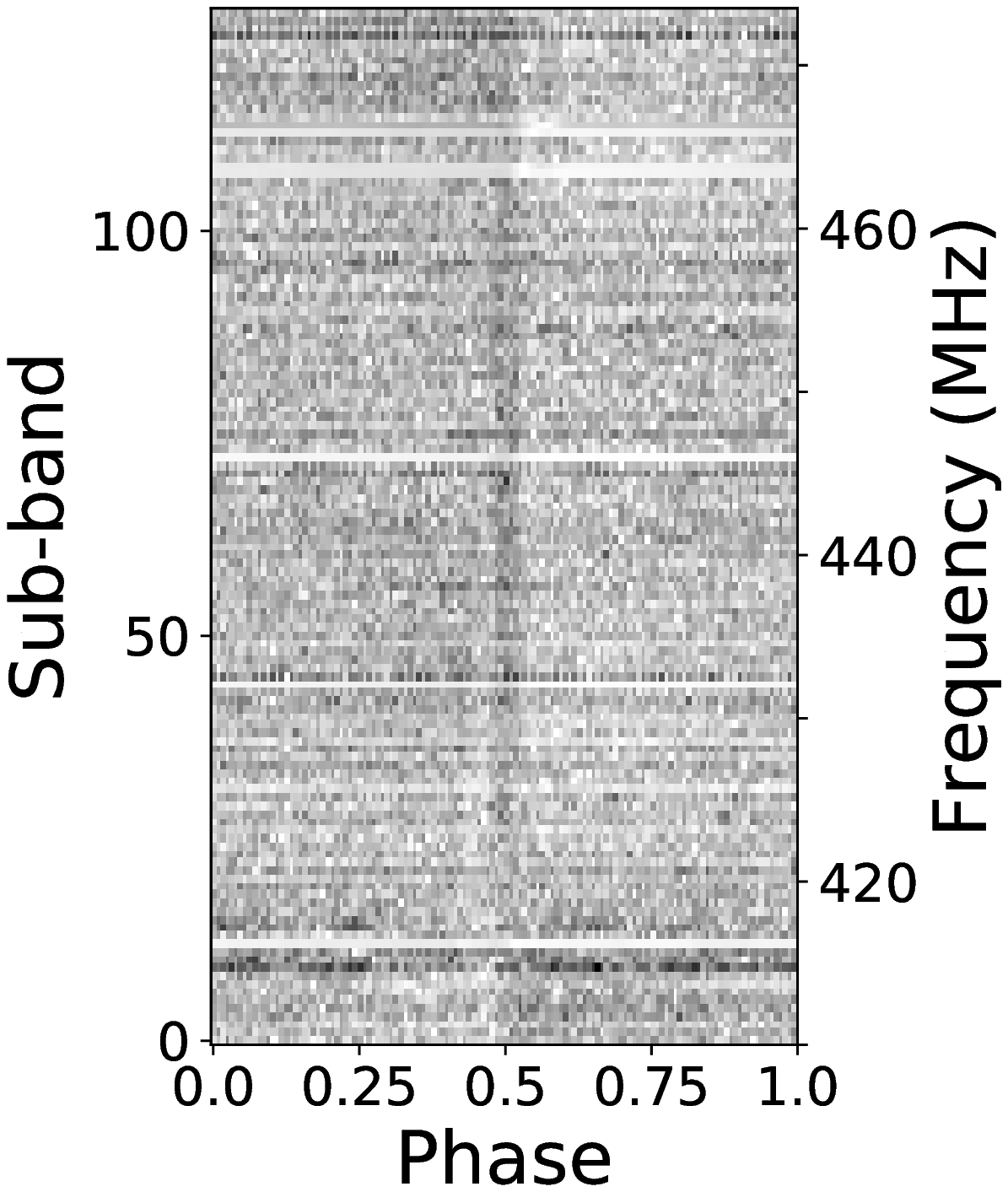}
    \end{minipage}
    \label{fig:wrong2}
    }
    \vspace{3ex}

    \subfloat[Missing pulsar 3]{
	\begin{minipage}[b]{0.5\linewidth}
    \centering
    \includegraphics[width=0.4\linewidth]{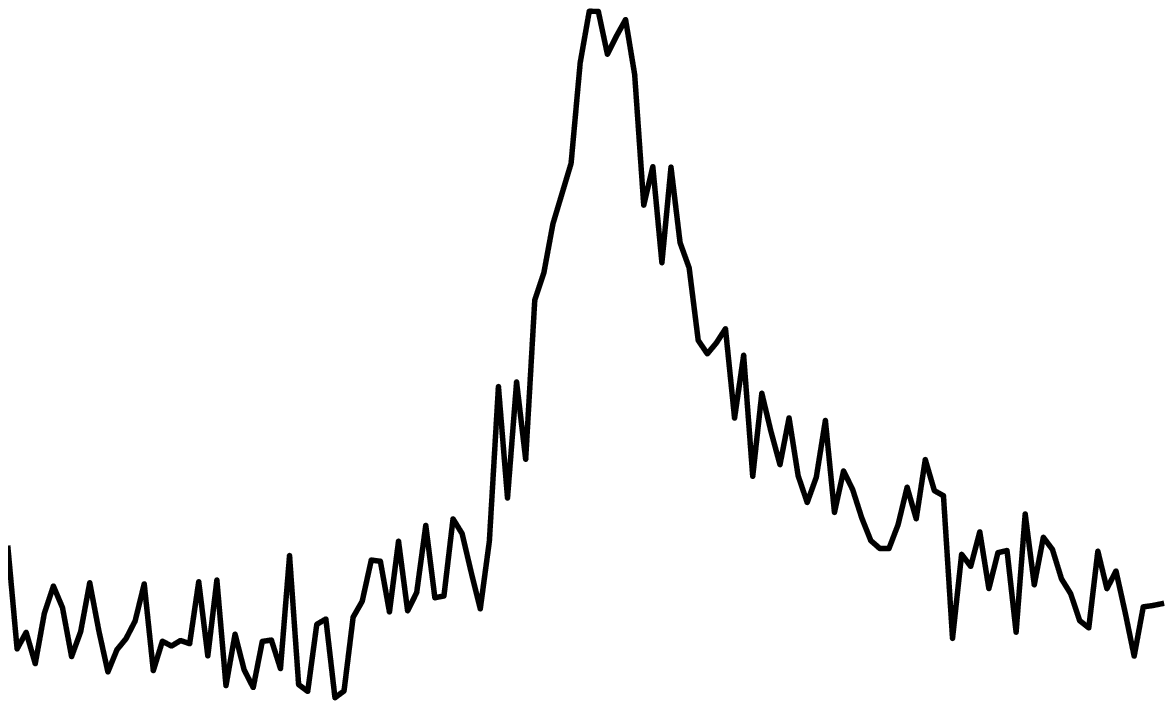}
    \includegraphics[width=0.4\linewidth]{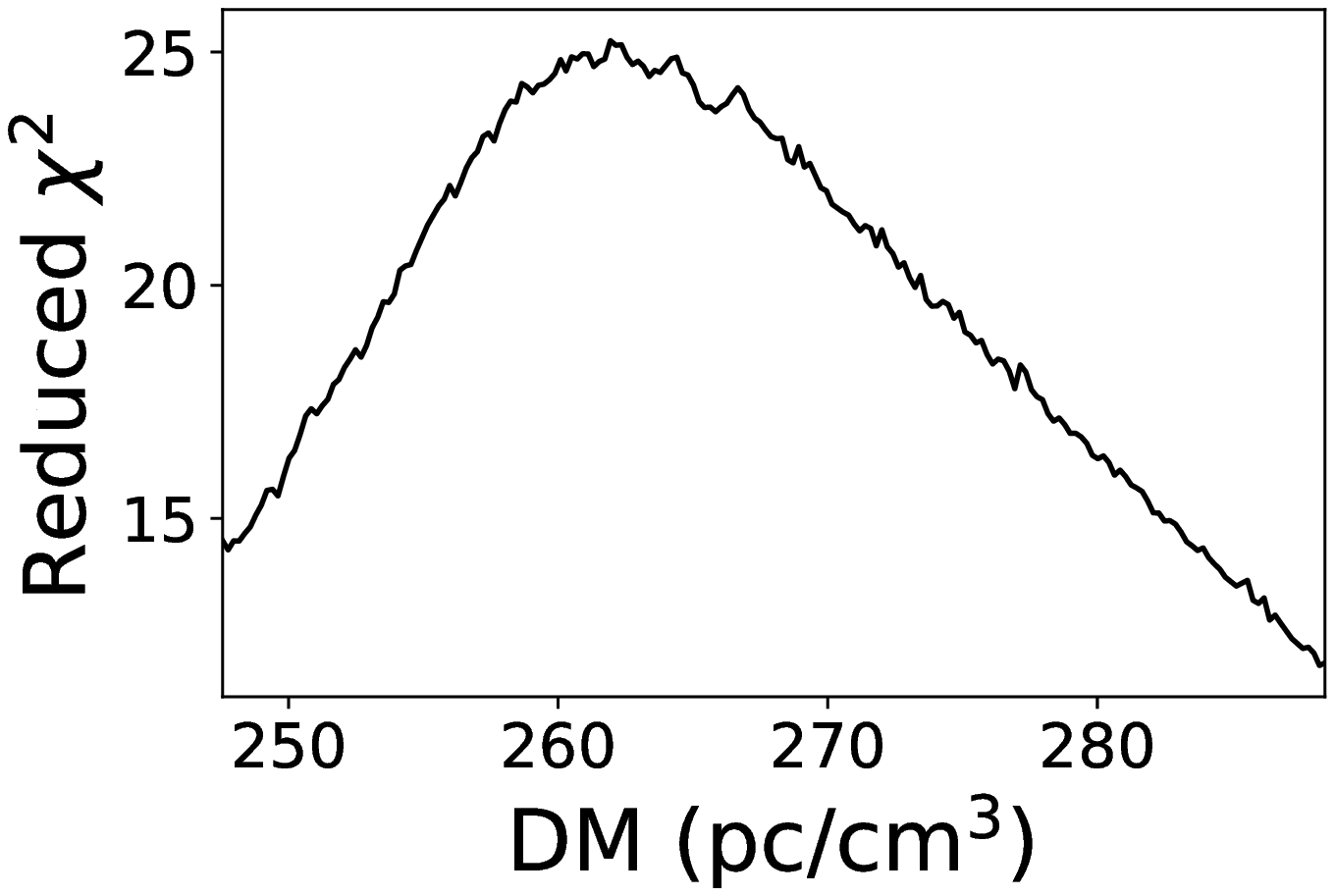}
    \vfill

    \includegraphics[width=0.4\linewidth]{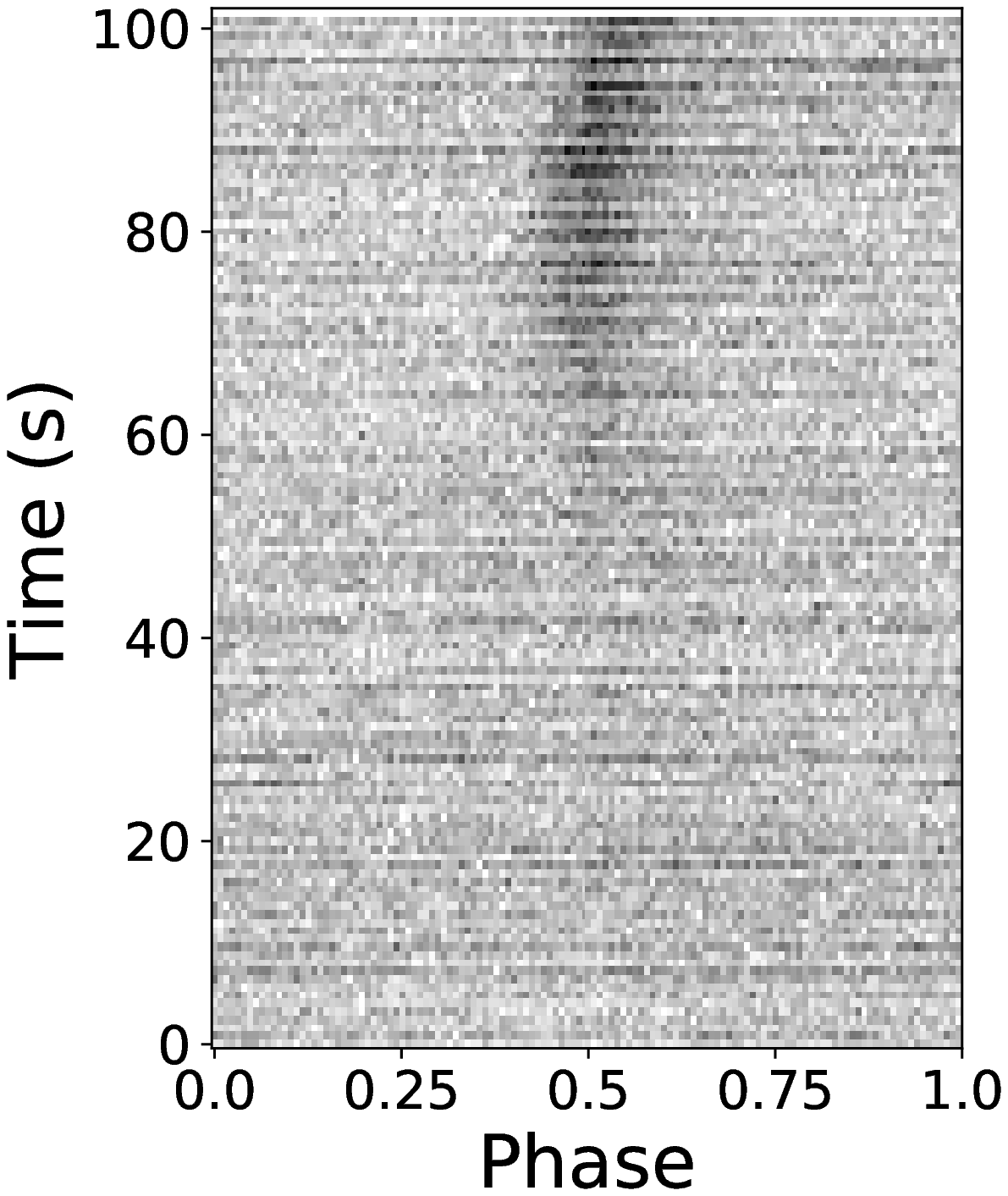}
    \includegraphics[width=0.4\linewidth]{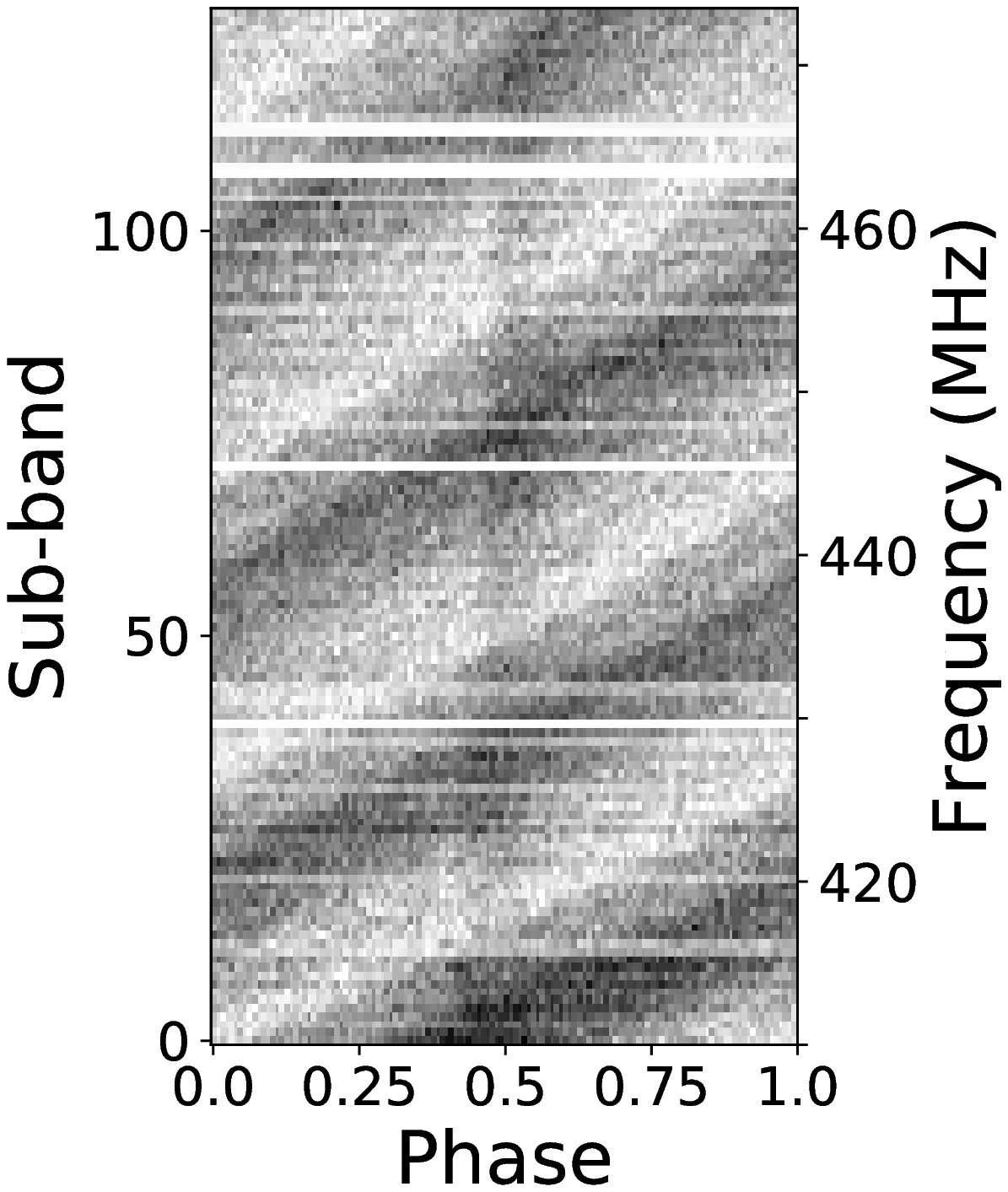}
    \end{minipage}
    \label{fig:wrong3}
    }
    \subfloat[Missing pulsar 4]{
	\begin{minipage}[b]{0.5\linewidth}
    \centering
    \includegraphics[width=0.4\linewidth]{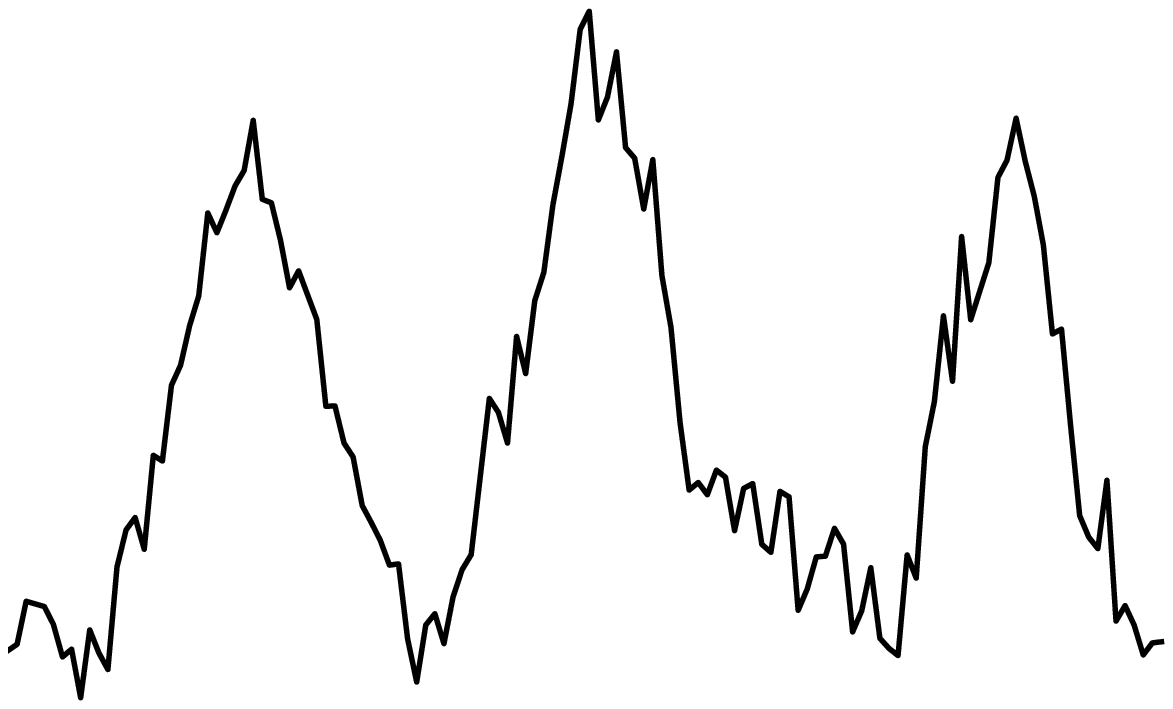}
    \includegraphics[width=0.4\linewidth]{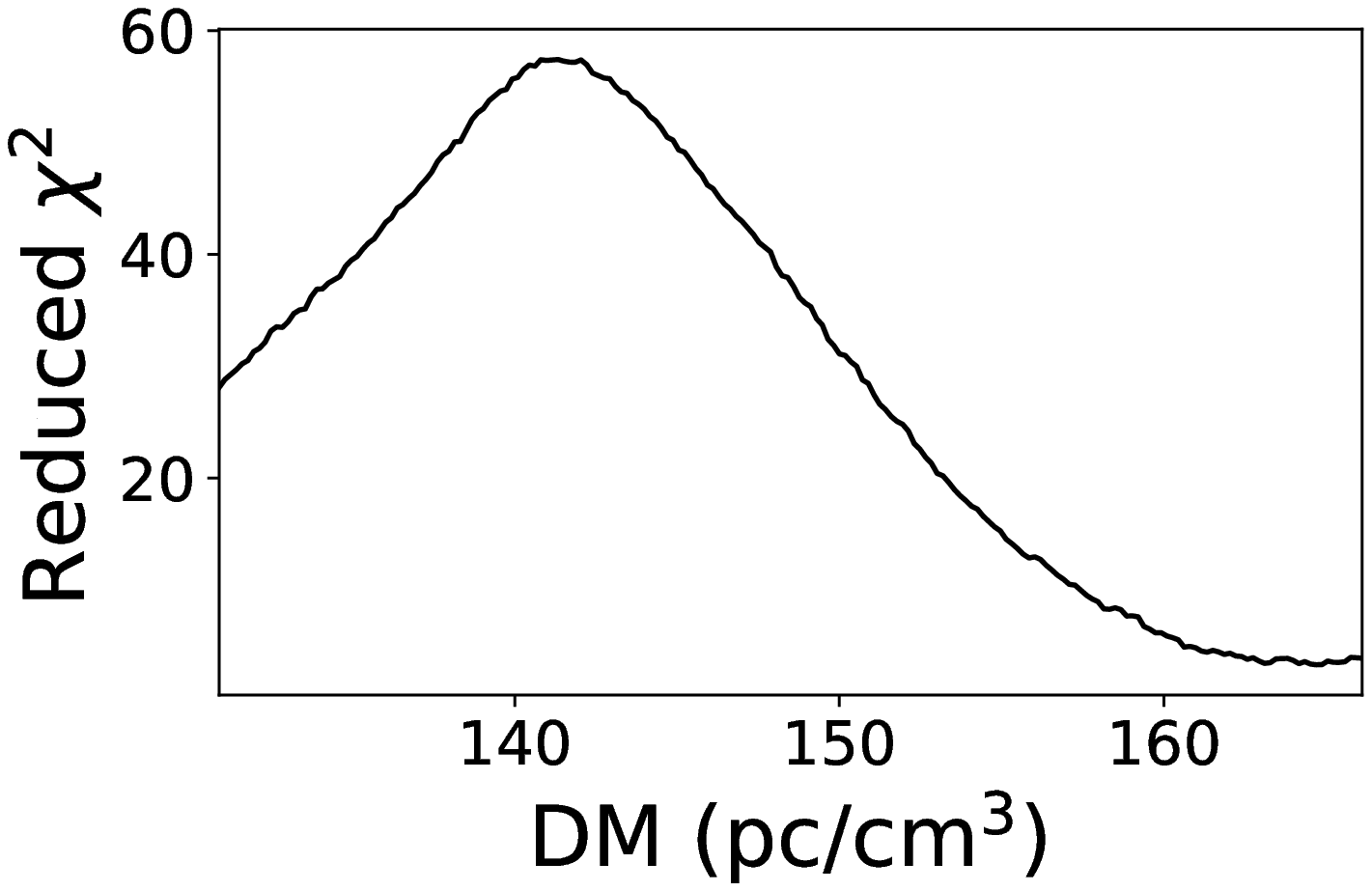}
    \vfill

    \includegraphics[width=0.4\linewidth]{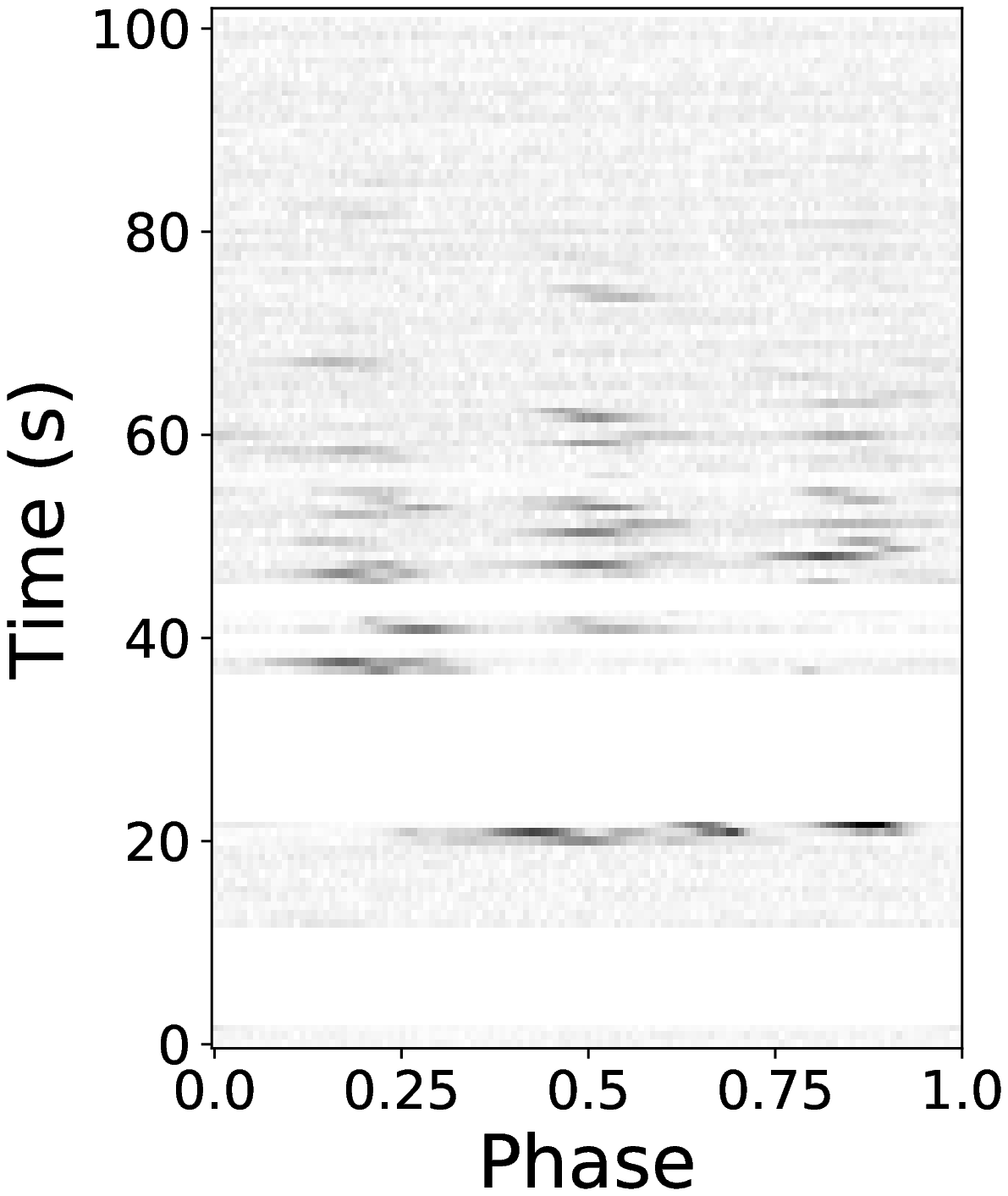}
    \includegraphics[width=0.4\linewidth]{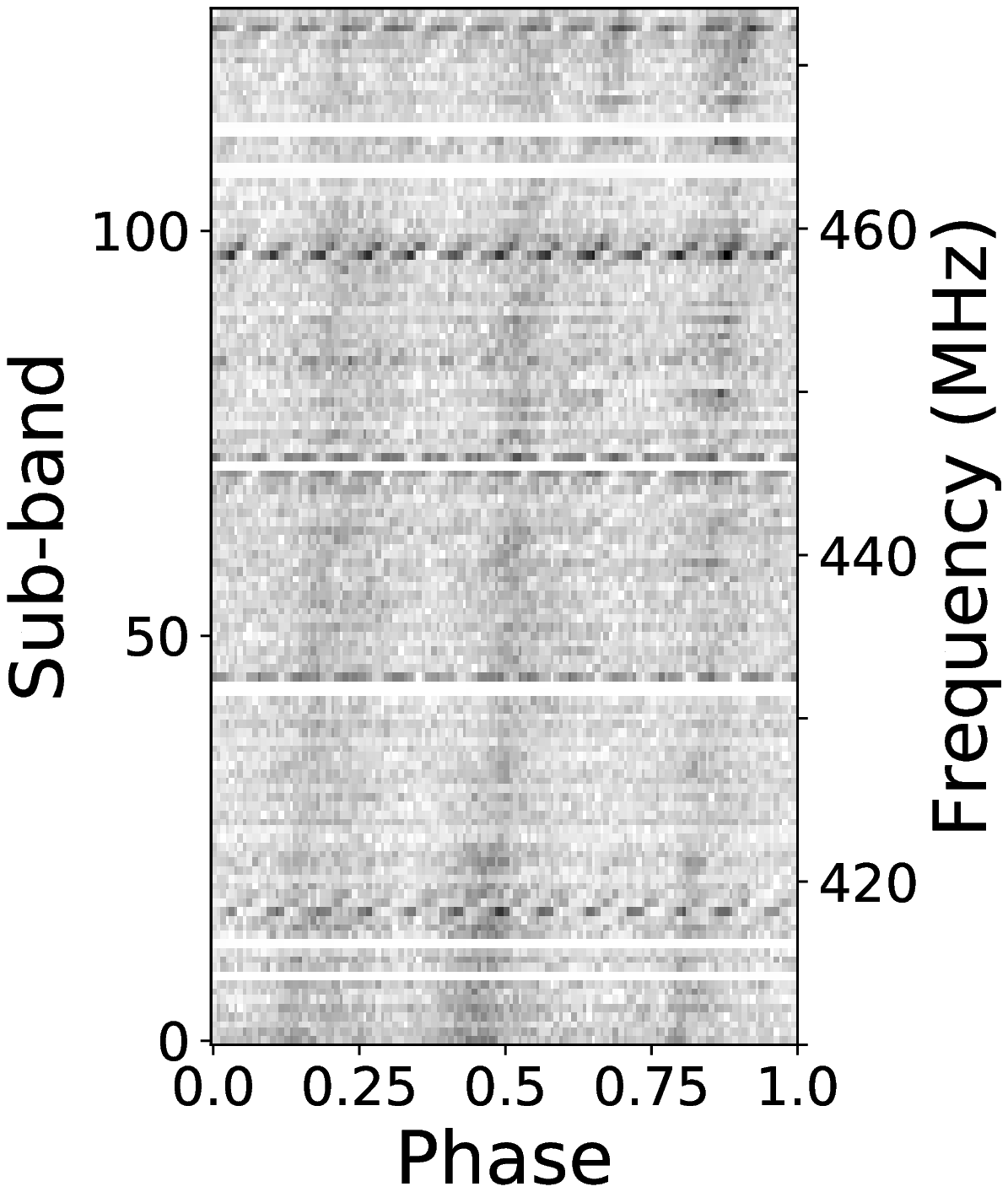}
    \end{minipage}
    \label{fig:wrong4}
    }
	
    \caption{The diagnostic subplots of the missing pulsars. There are four diagnostic subplots for each pulsar, the upper left one is the profile, the upper right one is DM curve, the bottom left one is the time versus phase plot and the bottom right one is the frequency versus phase plot.}
    \label{fig:miss}
\end{figure*}

Despite the prominent performance of CCNN on the FAST data, the missing pulsars of it are still required to be concerned about. In general, CCNN (V-CCNN) mistakenly classified 4 pulsars as the RFI. The diagnostic subplots of them are presented in Fig.~\ref{fig:miss} which guides us to detailedly analyse the characteristics of the missing pulsars.

In summary, there are relatively obvious ``pulsar-like'' patterns in the diagnostic subplots of all the missing pulsars. However, there is different confusing information in the subplots at the same time. They are summarized as follows.

\begin{itemize}
    \item \textbf{The presence of the RFI}: For these missing pulsars, the RFI can mainly be observed in the time versus phase plot and frequency versus phase plot. The strong and persistent one (e.g. the time versus phase plot in Fig.~\ref{fig:wrong2}) almost overshadows the pulsar signals. The periodic interference causes the oblique lines in the time versus phase plot (Fig.~\ref{fig:wrong1}) and the oblique lines in the frequency versus phase plot are caused by the RFI with zero DM (Fig.~\ref{fig:wrong3}). They are all likely to make some negative effects for CCNN to extract the pulsar signals.
    \item \textbf{Intensity variation}: The signal intensity varies over time and in some normalized subplots, the signals look like to be disappeared sometimes (e.g. the time versus phase plot in Fig.~\ref{fig:wrong3} and Fig.\ref{fig:wrong4}). This phenomenon is caused by the rotation of the beam pattern during the observation.
\end{itemize}


\section{Conclusions}
A novel model for pulsar candidate selection, CCNN, is proposed in this paper.
Different from the existing ML models for the pulsar candidate selection, CCNN extracts the features from four diagnostic subplots all utilizing the CNN rather than traditional ML algorithms. Besides, concatenate layer in CCNN has the ability to automatically learn the combination of the features extracted from the subplots. The introduction of this layer avoids labeling every subplot manually, which significantly reduces labor and expense. Finally, Experiments of the CCNN on the FAST data compared to state-of-the-art models, CCNN outperforms them in summary.

Although CCNN has achieved outstanding results on the FAST data, several improvements and extensions still can be made in future work in theory, to make the model more useful and powerful, as follows:
\begin{itemize}
    \item \textbf{Data collection and labeling}: All the classifiers mentioned in this paper tend to mistakenly identify some non-pulsars as the pulsar signal so that they all
    suffer from a relatively low precision which increases the labor and expense for the further observations. Collecting additional labeled data for training is an effective way to improve the performance of CCNN for pulsar candidate selection on the FAST data.
    \item \textbf{Abandon resizing the input data}: Data resizing includes interpolation for the small sized data and scrunching for the large sized data. As a result, the former operation is likely to bring about the useless or even wrong information into the original diagnostic subplots while the latter may discard some important patterns within the subplots, leading to inaccurate identification for the pulsar candidates. Actually, the global pooling makes the CCNN have the ability to accept the input data with arbitrary size in theory. This characteristic of CCNN will be detailedly discussed in our next work.
    \item \textbf{Robust learning}: Form the analysis of the missing pulsar, we can conclude that the ``pulsar-like'' patterns all appear in the diagnostic subplots but are overshadowed by the interference to varying degrees. In ML, this type of data is referred to as noisy data \citep{zhu2004class}. Robust learning is an effective way to address this problem since it has the ability to mitigate the negative effects coming from the noise. Therefore, robust learning methods can be added to the data preprocessing or the construction of models.
\end{itemize}

\section*{Acknowledgements}
This work  is supported by the National Natural Science Foundation of China (grant Nos 11973022, U1811464), the Natural Science Foundation of Guangdong Province (2020A1515010710) and the Joint Research Fund in Astronomy (U1531242) under cooperative agreement between the National Natural Science Foundation of China (NSFC) and Chinese Academy of Sciences (CAS).




\bibliographystyle{mnras}
\bibliography{example} 



%
%
%


\bsp	
\label{lastpage}
\end{document}